\documentclass[journal]{IEEEtran}

\usepackage{epsfig,amsmath,amssymb,epsf,amsthm,scalefnt,multirow,subfig}
\usepackage{xcolor}
\usepackage{float}
\usepackage{cite}
\usepackage{psfrag}
\usepackage{graphics,amsmath,amssymb,graphicx,amsthm,scalefnt,multirow,dsfont,subfig}

\usepackage{epsfig,amsmath,amssymb,epsf,amsthm,scalefnt,multirow}
\usepackage{xcolor}
\usepackage{float}
\usepackage{cite}
\usepackage{psfrag}
\usepackage{cleveref}
\usepackage{mathrsfs}
\usepackage{mathtools}
\usepackage{algpseudocode}
\usepackage{algorithm}
\usepackage{enumerate}
\usepackage{stfloats}
\usepackage{amsmath}

\usepackage{amsmath, amssymb}
\usepackage{array}
\usepackage{booktabs}
\usepackage{multirow}
\usepackage{tabularx}

\newtheorem{lemma}{Lemma}
\newtheorem*{lemma*}{Lemma}

\newtheorem{corollary}{Corollary}

\newtheorem{proposition}{Proposition}

  \def\cC{{\mathcal{C}}}

 \def\cN{{\mathcal{N}}} \def\cO{{\mathcal{O}}}

\def\diag{\mathop{\mathrm{diag}}}

\def\trace{\mathop{\mathrm{tr}}}

\def\Re{\mathop{\mathrm{Re}}}
\def\Im{\mathop{\mathrm{Im}}}

\def\bSigma{{\pmb{\Sigma}}} 
 
\def\bGamma{{\pmb{\Gamma}}} 
\def\bOmega{{\pmb{\Omega}}} 
 \def\btheta{{\pmb{\theta}}}
 
 \def\bmu{{\pmb{\mu}}}
\def\b0{{\pmb{0}}}\def\bLambda{{\pmb{\Lambda}}} 

\def\ba{{\mathbf{a}}}  \def\bc{{\mathbf{c}}} 
  \def\bg{{\mathbf{g}}} 
   
 \def\bn{{\mathbf{n}}}  
 \def\br{{\mathbf{r}}} \def\bs{{\mathbf{s}}} 
\def\bu{{\mathbf{u}}}   \def\bx{{\mathbf{x}}}
\def\by{{\mathbf{y}}} \def\bz{{\mathbf{z}}}  

\def\bA{{\mathbf{A}}} \def\bB{{\mathbf{B}}} \def\bC{{\mathbf{C}}} \def\bD{{\mathbf{D}}}
 \def\bF{{\mathbf{F}}} \def\bG{{\mathbf{G}}} 
\def\bI{{\mathbf{I}}}  \def\bK{{\mathbf{K}}} 
\def\bM{{\mathbf{M}}} \def\bN{{\mathbf{N}}}  
 \def\bR{{\mathbf{R}}} \def\bS{{\mathbf{S}}} 
\def\bU{{\mathbf{U}}} \def\bV{{\mathbf{V}}} \def\bW{{\mathbf{W}}} \def\bX{{\mathbf{X}}}
\def\bY{{\mathbf{Y}}} \def\bZ{{\mathbf{Z}}}

\begin{document}	
\bstctlcite{IEEEexample:BSTcontrol}
\title{Task-Based Quantization for Channel Estimation \\ in RIS Empowered MmWave Systems}

\author{\IEEEauthorblockN{Gyoseung Lee, In-soo Kim, Yonina C. Eldar, A. Lee Swindlehurst, Hyeongtaek Lee, Minje Kim,
		and Junil Choi}
	\thanks{This work was supported in part by Institute of Information \& communications Technology Planning \& Evaluation (IITP) under 6G·Cloud Research and Education Open Hub (IITP-2025-RS-2024-00428780) grant funded by the Korea government (MSIT),  by IITP grant funded by the Korea government (MSIT) (No.2021-000269, Development of sub-THz band wireless transmission and access core technology for 6G Tbps data rate), and by the {U.S.} National Science Foundation under grants CNS-2107182 and CCF-2225575. \textit{(Junil Choi and A. Lee Swindlehurst are the corresponding authors of this article.)}}
	\thanks{Gyoseung Lee, Minje Kim, and Junil Choi are with the School of Electrical Engineering, Korea Advanced Institute of Science and Technology, Daejeon 34141, South Korea (e-mail: \{iee4432; mjkim97; junil\}@kaist.ac.kr).}
	\thanks{In-soo Kim is with Wireless Research \& Development (WRD), Qualcomm Technologies, Inc., San Diego, CA, USA (e-mail: insookim@qti.qualcomm.com).}
	\thanks{Yonina C. Eldar is with the Faculty of Math and Computer Science, Weizmann Institute of Science, Rehovot, Israel (e-mail: yonina.eldar@weizmann.ac.il)}
	\thanks{A. Lee Swindlehurst is with the Center for Pervasive Communications and Computing, Henry Samueli School of Engineering, University of California, Irvine, CA 92697, USA (e-mail: swindle@uci.edu).}
	\thanks{Hyeongtaek Lee is with the Department of Electronic and Electrical Engineering, Ewha Womans University, Seoul 03760, South Korea (e-mail: htlee@ewha.ac.kr).}
	}
\maketitle

\begin{abstract}
In this paper, we investigate channel estimation for reconfigurable intelligent surface (RIS) empowered millimeter-wave (mmWave) multi-user single-input multiple-output communication systems using low-resolution quantization.
Due to the high cost and power consumption of analog-to-digital converters (ADCs) in large antenna arrays and for wide signal bandwidths, designing mmWave systems with low-resolution ADCs is beneficial.
To tackle this issue, we propose a channel estimation design using task-based quantization that considers the underlying hybrid analog and digital architecture in order to improve the system performance under finite bit-resolution constraints.
Our goal is to accomplish a channel estimation task that minimizes the mean squared error distortion between the true and estimated channel. We develop two types of channel estimators: a cascaded channel estimator for an RIS with purely passive elements, and an estimator for the separate RIS-related channels that leverages additional information from a few semi-passive elements at the RIS capable of processing the received signals with radio frequency chains.
Numerical results demonstrate that the proposed channel estimation 
designs exploiting task-based quantization outperform purely digital methods and can effectively approach the performance of a system with unlimited resolution ADCs. 
Furthermore, the proposed channel estimators are shown to be superior to baselines with small training overhead.
\end{abstract}


\section{Introduction}\label{sec1}
In millimeter-wave (mmWave) communication systems operating at high carrier frequencies ranging from 30-300 GHz, the availability of large bandwidths allows for high data rates, thereby supporting various requirements anticipated for future wireless communications \cite{Wang:2018}.
However, high frequency propagation leads to significant path-loss and strong directivity, making the signals vulnerable to blockages.
To tackle these issues,metasurface-based technologies, such as reconfigurable intelligent surfaces (RISs), which are planar metamaterial structures consisting of low-cost passive scattering elements \cite{Wu:2020, Renzo:2019, Basar:2019, Renzo:2022}, and reconfigurable holographic surfaces (RHSs), which operate as reconfigurable antenna arrays at a base station (BS) with a reduced number of radio frequency (RF) chains to enhance cost and energy efficiency \cite{HMIMO_1, HMIMO_2}, have emerged as promising solutions for beyond-5G communications.
In the case of RIS, the propagation environment can be adaptively adjusted in beneficial ways by intelligently modifying its reflection properties \cite{Wu:2020, Renzo:2019, Basar:2019, Renzo:2022}. For instance, an RIS can effectively construct a virtual line-of-sight link between a BS and user equipment (UE) that would otherwise be blocked, thus enhancing mmWave system coverage despite the presence of obstacles \cite{Wu:2020, Basar:2019}.

To fully leverage the benefits of an RIS, it is necessary to acquire accurate channel state information (CSI) at the BS or UE. However, an RIS typically lacks baseband processing capability as it comprises purely passive elements without RF chains, making CSI acquisition for the RIS-related channels inherently difficult. To tackle this challenge, most existing works  propose estimating the cascaded UE-RIS-BS channel, which is generally sufficient when designing the RIS phase shifts for data transmission \cite{Swindlehurst:2022, Pan:2022, Kim:2022, Wei:2021}.
However, in some applications such as those involving UE localization, obtaining CSI for the individual BS-RIS and RIS-UE channels is necessary \cite{Guo:2020, Zhang:2020, Hua:2023}. To address this issue, some existing works have suggested integrating the RIS with a few semi-passive elements equipped with their own RF chain and analog-to-digital converters (ADCs), enabling the reception and processing of training~signals \cite{Taha:2021, Liu:2020, Jin:2021, Zhang:2021, Gyoseung:2023}.

Prior work on channel estimation for both types of RIS has mostly assumed infinite resolution ADCs at the BS or RIS. In practical applications, developing systems that operate with low-resolution ADCs helps mitigate the overall cost and power consumption, particularly in mmWave massive multiple-input multiple-output (MIMO) systems that employ large antenna arrays and high bandwidths. Prior work that has considered low-resolution ADCs for this problem includes \cite{Wang:2023}, which developed a bilinear generalized approximate message passing (BiG-AMP) approach for cascaded channel estimation in single-user multiple-input single-output systems. 
In \cite{Han:2025}, a leakage structure orthogonal matching pursuit (LS-OMP) algorithm was proposed to estimate the cascaded channel taking into account leakage caused by grid mismatch.
Also, \cite{In-soo:2023} proposed a Bayesian estimator to estimate the separate RIS channels for multi-user single-input multiple-output (MU-SIMO) systems, exploiting additional information from semi-passive elements equipped with low-resolution ADCs.
In \cite{Cao:2024}, a generalized multiple measurement vector (GMMV) problem and matrix completion were used to estimate the individual RIS channels, and a hierarchical message passing-based algorithm was proposed based on AMP-like approximations.
However, in mmWave systems, deploying large antenna arrays at the BS still results in significant overhead in terms of the total number of quantized bits that must be processed during the channel estimation phase. Our goal in this work is to design a channel estimator using low-resolution ADCs that achieves high estimation accuracy with a minimal number of quantization bits at the receiver.

Among the various quantization strategies that have been proposed, \cite{Shlezinger:2019, Shlezinger:2019_2, Shlezinger:2020, Salamatoan:2019, Bernardo:2023} have shown that co-designing the quantizer together with a hybrid analog and digital architecture can dramatically improve the performance. In conventional ``task-ignorant'' systems where the quantizer is designed solely to discretize the measurements, the desired task (recovering the transmitted signal vector) is performed separately in the digital domain.
On the other hand, {\em task-based quantization} captures lower-dimensional information from signals in the analog domain based on the specific system task, allowing for a reduction in the overall number of bits required to accomplish the task, thereby minimizing the power consumption and memory requirements.
Task-based quantization has been applied in various applications such as channel estimation \cite{Shlezinger:2019, Shlezinger:2019_2, Bernardo:2023}, quadratic function recovery \cite{Salamatoan:2019}, MIMO radar \cite{Xi:2021}, and dual functional radar and communication systems \cite{Ma:2021}. However, prior work has not addressed task-based quantization in RIS-aided communication systems, except for our preliminary work in \cite{Conference_version} which only addressed cascaded channel~estimation.

In this paper, motivated by the success of task-based quantization over conventional all-digital approaches independent of the task \cite{Shlezinger:2019, Shlezinger:2019_2, Shlezinger:2020, Salamatoan:2019, Bernardo:2023}, we apply task-based quantization to the problem of channel estimation for RIS-empowered MU-SIMO systems, where identical scalar ADCs are used to mitigate hardware cost and energy limitations.
Specifically, we consider two approaches: 1) cascaded channel estimation for cases involving an RIS with purely passive elements, and 2) separate estimation of the RIS channels by exploiting an RIS with a few semi-passive elements connected with low-resolution ADCs that provide additional local information.
Note that deploying low-resolution ADCs for the semi-passive elements of an RIS is consistent with the goal that such surfaces operate with minimal additional cost and power consumption.
For both channel estimation approaches, the goal is to minimize the mean squared error (MSE) in estimating the channel for a hybrid analog and digital architecture by proper choice of the analog and digital combining matrices and the ADC quantization range or support.
Our main contributions are as follows:
\begin{itemize}
	\item For an RIS equipped with purely passive elements, we first propose a cascaded channel estimation framework using task-based quantization to minimize the total number of quantization bits at the BS while employing low-resolution ADCs. Combining the mmWave channel structure with task-based quantization, we demonstrate that, under typical mmWave system conditions, the dimensionality of the BS observations in the analog domain can be significantly reduced, depending on the number of propagation paths that comprise the RIS-related channels. Due to the limited RF scattering in mmWave bands, the number of propagation paths is limited, which enables the proposed channel estimator to achieve a significant reduction in the total number of quantization bits at the BS compared to task-ignorant approaches.
	\item For an RIS with semi-passive elements, we propose a two-stage estimator of the RIS channel components using task-based quantization that first estimates the UE-RIS channels based on observations from the semi-passive elements of the RIS, and subsequently estimates the BS-RIS channel using the UE-RIS channel estimate and observations at the BS. This estimator is the first to consider low-resolution quantization at both the BS and the semi-passive elements of the RIS. We also demonstrate that the total number of quantization bits required to estimate the individual channel components at the BS can be further reduced compared to the proposed cascaded channel estimator.
	\item Our numerical results verify that, in terms of channel estimation accuracy, the proposed cascaded channel estimator closely approaches the performance of the minimum MSE (MMSE) estimator without quantization \cite{MMSE:book}, and outperforms a purely digital approach assuming identical bit-resolution ADCs in which a task-ignorant MMSE estimator based on \cite{Shlezinger:2019} is applied to the quantized measurements.
	Subsequently, we demonstrate that the proposed channel estimator for the separate RIS channels shows performance comparable to an MMSE estimator without quantization even with a small number of semi-passive elements.
	Finally, we show that the proposed channel estimators outperform baseline approaches including the methods of \cite{Wang:2023, In-soo:2023} that use low-resolution ADCs without task-based quantization, achieving a significant reduction in the total number of quantization bits and a small training overhead.
\end{itemize}

The rest of the paper is organized as follows.
Section \ref{sec2} presents the system model for the assumed RIS-aided MU-SIMO system.
In Section \ref{sec3}, the problem formulations for obtaining the cascaded and separate channel estimates are investigated.
The basic concept of task-based quantization is explained in Section \ref{sec4}, and the proposed channel estimators based on task-based quantization are developed in Sections \ref{sec5} and \ref{sec6}.
Numerical results for the proposed approaches are provided in Section \ref{sec_numerical}, and we conclude the paper in Section \ref{conclusion}.

$Notation$: Lower- and upper-case boldface letters represent column vectors and matrices, respectively.
The conjugate, transpose, and conjugate transpose of a matrix $\bA$ are denoted by $\bA^*$, $\bA^{\mathrm{T}}$, and $\bA^{\mathrm{H}}$, respectively.
For a square matrix $\bA$, $\trace(\bA)$ and $\bA^{-1}$ are respectively the trace and inverse of $\bA$.
The quantities $[\bA]_{i,:}$ and $[\bA]_{i,j}$ denote the $i$-th row and the $(i,j)$-th entry of matrix $\bA$, respectively. The operator
$\mathrm{vec}(\bA)$ denotes the vectorization of matrix $\bA$, and $\mathrm{unvec}(\ba)$ denotes the inverse operation.
The notation $\mathrm{diag}(\ba)$ represents a diagonal matrix whose diagonal elements correspond to the entries of the vector $\ba$, and $\mathrm{blkdiag}(\cdot)$ denotes a block-diagonal matrix with blocks defined by the argument.
The $\ell_p$-norm of a vector $\ba$ and the Frobenius-norm of a matrix $\bA$ are respectively denoted by $\Vert \ba \Vert_p$ and $\Vert \bA \Vert_{\mathrm{F}}$.
A circularly symmetric complex Gaussian distribution with mean vector $\bmu$ and covariance matrix $\bK$ is represented using $\cC\cN(\bmu,\bK)$. 
The quantities $\b0_{m}$, $\boldsymbol{1}_{m}$, and $\bI_m$ represent the $m \times 1$ all-zeros vector, the $m \times 1$ all-ones vector, and the $m \times m$ identity matrix, respectively. 
The expressions $\vert a \vert$, $\angle a$, $\Re(a)$, and $\Im(a)$ represent the magnitude, angle, real part, and imaginary part of a complex number $a$, respectively.
The expression $a^+$ denotes $\max(a, 0)$ for a real number $a$.
The Hadamard product, Kronecker product, and Khatri-Rao product of matrices $\bA$ and $\bB$ are denoted as $\bA \odot \bB$, $\bA \otimes \bB$, and $\bA \diamond \bB$, respectively.
For convenience, we summarize the notation that will be used throughout the paper in Table \ref{Table_notation}.
\begin{table}
	\caption{Summary of notation used in the paper.}
	\label{Table_notation}
	\centering
	\renewcommand{\arraystretch}{1.2}
	\begin{tabular}{@{}c p{6.5cm}@{}}
		\toprule
		\textbf{Notation} & \textbf{Description} \\
		\midrule
		$N$                     & Number of BS antennas \\ \hline
		$L=L_{\mathrm{h}}L_{\mathrm{v}}$              & Number of total RIS elements \\  \hline
		$L_{\mathrm{h}}, L_{\mathrm{v}}$                  & Number of horizontal and vertical elements of RIS  \\  \hline
		$L_{\mathrm{p}}, L_{\mathrm{a}}$              & Number of passive and semi-passive RIS elements \\  \hline
		$K$                  & Number of UEs \\  \hline
		$T_{\mathrm{p}}=T\tau$                  & Number of total time slots for pilot training \\  \hline
		$T$                  & Number of subblocks in $T_{\mathrm{p}}$ \\  \hline
		$\tau$                  & Number of time slots in each subblock\\  \hline
		$P_k$              & Transmit power at the $k$-th UE \\  \hline
		$M_{\mathrm{RB}}$              & Number of propagation paths in the RIS-BS channel \\  \hline
		$M_{\mathrm{UR},k}$              & Number of propagation paths in the $k$-th UE-RIS channel \\  \hline
		$\bG$              & Uplink channel from the RIS to the BS \\  \hline
		$\mathbf{f}_k$              & Uplink channel from the $k$-th UE to the RIS \\  \hline
		$\bF$              & Uplink channel from the UEs to the RIS \\  \hline
		$\bC$              & Cascaded channel between the BS and the UEs \\  \hline
		$\btheta[t]$                   & Vector of RIS reflection coefficients \\ \hline
		$\bOmega$, $\bOmega^{\mathrm{c}}$                   & Index vectors of semi-passive and passive RIS elements \\ \hline
		$\bs[t]$                   & Vector of passive RIS reflection coefficients \\ \hline
		$\by[t], \bz[t]$ & Received signals at the BS and semi-passive elements\\ \hline
		$\boldsymbol{\pi}_{\bz}$ & Quantized signals at the semi-passive RIS elements \\ \hline
		$x_k[t]$ & Transmit signal from the $k$-th UE \\ \hline
		$\bn_{\mathrm{B}}[t], \bn_{\mathrm{R}}[t]$ & AWGN noise at the BS and RIS \\ \hline
		$Q_{\tilde{\nu}}(\cdot)$ & Scalar ADC with resolution $\log_2 \tilde{\nu}$ bits \\ \hline
		$\gamma_{\bc}, \gamma_{\bg}$ & ADC thresholds for $\bC$ and $\bG$ \\ \hline
		$\bB_{\bc}, \bB_{\bg}$ & Analog combining matrices for $\bC$ and $\bG$ \\ \hline
		$\bD_{\bc}, \bD_{\mathbf{f}}, \bD_{\bg}$ & Digital processing matrices for $\bC$, $\bF$, and $\bG$ \\ \hline
		$G_{\bc}, G_{\bg}$ & Number of scalar ADCs for $\bC$ and $\bG$ \\ \hline
		$\boldsymbol{\pi}_{\bc}, \boldsymbol{\pi}_{\bg}$ & Outputs of scalar ADCs for $\bC$ and $\bG$ \\
		\bottomrule
	\end{tabular}
	\vspace{-5mm}
\end{table}

\section{System Model}\label{sec2}

\subsection{Signal model}
\begin{figure}
	\centering
	\includegraphics[width=1.0\columnwidth]{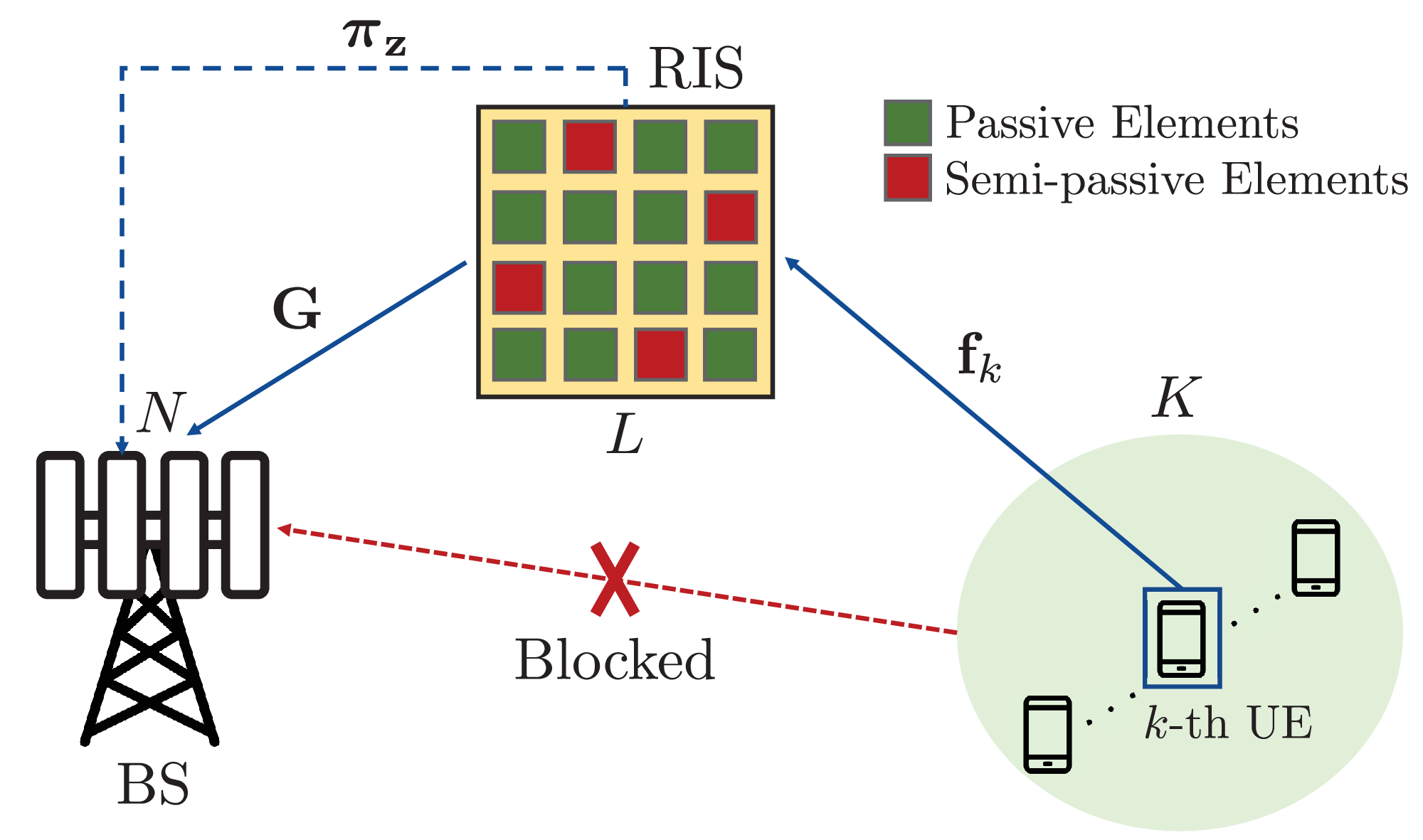}
	\caption{An example of an uplink RIS-aided mmWave MU-SIMO system with semi-passive elements. There are $N$ BS antennas, $K$ UEs with a single antenna, and $L$ total RIS~elements.}	\label{system_model}
\end{figure}
We investigate an uplink RIS-aided MU-SIMO scenario as illustrated in Fig. \ref{system_model}, where a BS with $N$ antennas in a uniform linear array (ULA) communicates with $K$ single-antenna UEs. We assume a mmWave setting with large $N$, so the BS adopts both hybrid analog and digital beamforming together with low-resolution ADCs to reduce the required cost and power consumption. A single RIS composed of a uniform planar array (UPA) with $L=L_{\mathrm{h}}L_{\mathrm{v}}$ elements is present, where $L_{\mathrm{h}}$ and $L_{\mathrm{v}}$ denote the number of horizontal rows and vertical columns, respectively. The configuration of the RIS is under the control of the BS via a low-rate network connection.
Among the $L$ RIS elements, $L_{\mathrm{p}}$ passive elements reflect the incoming signals, while $L_{\mathrm{a}}=L-L_{\mathrm{p}} \ll L$ semi-passive elements operate in sensing mode during the channel estimation phase and in reflecting mode during the data transmission phase\footnote{A semi-passive RIS element is equipped with a sensor implemented with a low cost receive RF chain, allowing it to process the received signal \cite{Zhang:2021, Wu:2021}. In the sensing mode, the reflecting element is turned off, and the sensor is activated to receive pilot signals. In the reflection mode, the sensor is deactivated, and the reflecting element is turned on, operating like a conventional passive RIS element.}. 
Since the RIS is intended to be a low-power device, the baseband outputs of the semi-passive elements are sampled with low-resolution ADCs \cite{Zhang:2021, In-soo:2023}. In general, the goal is to use as few semi-passive elements as possible to minimize the additional cost and power consumption at the RIS. 

As in \cite{Guo:2021,Wang:2021,Zhang:2021}, the channels between the BS and UEs are assumed to be entirely blocked. Based on a block fading channel model, we assume that the BS-RIS and RIS-UE channels are constant during $T_{\mathrm{p}}$ time slots.
To indicate how the RIS is partitioned into semi-passive and passive elements, let $\bOmega \in \{0,1\}^{L}$ and $\bOmega^{\mathrm{c}} = \boldsymbol{1}_L - \bOmega$ denote the index vectors of the semi-passive and the passive elements, respectively, such that $\Vert \bOmega \Vert_0 = L_{\mathrm{a}}$. Focusing on uplink transmissions, the signal received at the BS prior to the hybrid analog/digital combining is given~by
\begin{align} \label{y_t}
	\by[t] &= \sum_{k=1}^K \bG\diag(\underbrace{\bOmega^{\mathrm{c}} \odot \btheta[t]}_{=\bs[t]})\mathbf{f}_k x_k[t] + \bn_{\mathrm{B}}[t],
\end{align}
where $x_k[t] \in \mathbb{C}$ is the transmit signal from the $k$-th UE satisfying $\mathbb{E}[\vert x_k[t] \vert^2] \leq P_k$, and $\bn_{\mathrm{B}}[t] \sim \cC\cN(\b0_N, \sigma_{\mathrm{B}}^2 \bI_N)$ is additive white Gaussian noise (AWGN) at the BS with variance $\sigma_{\mathrm{B}}^2$. The uplink channel from the RIS to the BS is denoted by  $\bG \in \mathbb{C}^{N \times L}$, and the uplink channel from the $k$-th UE to the RIS is represented by $\mathbf{f}_k \in \mathbb{C}^{L \times 1}$. The vector of RIS reflection coefficients is $\btheta[t] = [\theta_1[t],\cdots,\theta_L[t]]^{\mathrm{T}} \in \mathbb{C}^{L \times 1}$, with reflection amplitudes $\vert \theta_{\ell}[t] \vert = 1$ and phase shifts $\angle \theta_{\ell}[t] \in [0, 2\pi)$. The model for the hybrid beamforming and low-resolution quantization of $\by[t]$ will be discussed in Sections \ref{sec4} and \ref{sec5}.
Note that, although we assume a narrowband channel model throughout this paper, the proposed channel estimators are also applicable without loss of generality to wideband systems employing orthogonal frequency division multiplexing (OFDM), since each subcarrier typically experiences a narrowband channel.

If the RIS is equipped with a few semi-passive elements, it is possible to estimate the individual RIS channels. In this case, the RIS sends the locally sampled information from these elements to the BS. 
The signal received at time $t$ by the semi-passive RIS elements prior to quantization is denoted by $\bz[t]=\bOmega \odot (\bF \bx[t] + \bn_{\mathrm{R}}[t])$, with $\bF=[\mathbf{f}_1,\cdots,\mathbf{f}_K] \in \mathbb{C}^{L \times K}$, $\bx[t]=[x_1[t],\cdots,x_K[t]]^{\mathrm{T}} \in \mathbb{C}^{K \times 1}$, and where $\bn_{\mathrm{R}}[t] \sim \cC \cN (\b0_L, \sigma_{\mathrm{R}}^2 \bI_L)$ represents AWGN at the RIS with variance $\sigma_{\mathrm{R}}^2$. Stacking these signals over the $T_{\mathrm{p}}$ time slots leads~to
\begin{align} \label{Z_original}
	\bZ &= [\bz[1],\cdots,\bz[T_{\mathrm{p}}]] \nonumber \\ &= \bar{\bOmega} \odot (\bF \bX_{\mathrm{I}} + \bN_{\mathrm{R}} ),
\end{align}
where $\bar{\bOmega}=\bOmega \otimes \boldsymbol{1}_{T_{\mathrm{p}}}^{\mathrm{T}} \in \mathbb{C}^{L \times T_{\mathrm{p}}}$, $\bX_{\mathrm{I}}=[\bx[1],\cdots,$ $\bx[T_{\mathrm{p}}]] \in \mathbb{C}^{K \times T_{\mathrm{p}}}$, and $\bN_{\mathrm{R}}=[\bn_{\mathrm{R}}[1],\cdots,\bn_{\mathrm{R}}[T_{\mathrm{p}}]] \in \mathbb{C}^{L \times T_{\mathrm{p}}}$. 
Both the real and imaginary parts of each element in $\bZ$ are quantized using identical low-resolution ADCs. Denoting the vectorized representation of (\ref{Z_original}) as $\bz=\mathrm{vec}(\bZ)$, the quantized version of $\bz$ is denoted by $\boldsymbol{\pi}_{\bz}$, which is forwarded to the BS. The $i$-th element of $\boldsymbol{\pi}_{\bz}$ is given by
\begin{align}
	\pi_{\bz,i}=
	\begin{cases}
		Q_{\tilde{\nu}}(z_i) \quad &\mbox{if } [\mathrm{vec}(\bar{\bOmega})]_i = 1 \\
		0 &\mbox{if } [\mathrm{vec}(\bar{\bOmega})]_i = 0,
	\end{cases}
\end{align}
where $Q_{\tilde{\nu}}(\cdot)$ denotes the ADC quantization operation with $\log_2 \tilde{\nu}$ bits of resolution, applied separately to the real and imaginary parts. The specific ADC model will be discussed in Section \ref{sec4_1}.

\textit{Remark:} As described in (\ref{y_t}) and (\ref{Z_original}), we consider a model in which the location of the semi-passive elements is fixed. This is different from the model in \cite{In-soo:2023, Cao:2024}, where a switching network between the RF chains and RIS elements is introduced to allow the selection of semi-passive elements to change at each time instant and improve channel estimation performance. However, such a network requires analog connections from the RF chains to all RIS elements via switches, which would be difficult to implement in practice, especially for a large RIS. As will be shown in Section \ref{sec_numerical_practical}, the proposed individual channel estimator based on task-based quantization without the switching network achieves superior channel estimation performance compared to \cite{In-soo:2023, Cao:2024}.

\subsection{Channel model}
To properly describe the mmWave channels, we adopt a geometric model \cite{Geometric_1, Geometric_2, Zhang:2021} due to the limited scattering environment in mmWave bands. The channel $\bG$ from the RIS to the BS is then represented by
\begin{align} \label{G}
	\bG &= \sqrt{\frac{NL}{M_{\mathrm{RB}}}} \sum_{m=1}^{M_{\mathrm{RB}}} \alpha_{\mathrm{RB},m} \ba_{\mathrm{B}}(\phi_{m}) \ba_{\mathrm{R}}^{\mathrm{H}}(\theta_{\mathrm{RB},m}^{\mathrm{Azi}}, \theta_{\mathrm{RB},m}^{\mathrm{Ele}}),
\end{align}
where $M_{\mathrm{RB}}$ denotes the number of propagation paths between the BS and RIS,  and $\alpha_{\mathrm{RB},m} \sim \cC\cN(0, \sigma_{\mathrm{RB}}^2)$ is the complex gain of the $m$-th path, which is independent and identically distributed (i.i.d.) with variance $\sigma_{\mathrm{RB}}^2$ that depends on the path-loss. The array steering vectors at the BS and RIS 
are expressed as $\ba_{\mathrm{B}}(\cdot) \in \mathbb{C}^{N\times 1}$ and $\ba_{\mathrm{R}}(\cdot) \in \mathbb{C}^{L \times 1}$, respectively, and $\ba_{\mathrm{B}}(\phi_{m})$ is represented by
\begin{align}
	\ba_{\mathrm{B}}(\phi_{m})=\frac{1}{\sqrt{N}}[1,e^{j\omega_{m}},\cdots, e^{j(N-1)\omega_m} ]^{\mathrm{T}},
\end{align}
where $\phi_{m}$ is the angle of arrival (AoA) of the $m$-th path and $\omega_m=2\pi d_{\mathrm{B}}\sin(\phi_{m})/\lambda_{\mathrm{c}}$ represents the spatial frequency with carrier wavelength $\lambda_{\mathrm{c}}$ and BS antenna spacing $d_{\mathrm{B}}$. The steering vector at the RIS is
\begin{align} \label{a_R}
	\ba_{\mathrm{R}}(\theta_{\mathrm{RB},m}^{\mathrm{Azi}}, \theta_{\mathrm{RB},m}^{\mathrm{Ele}}) = \ba_{\mathrm{R,v}}(\psi_{m}) \otimes \ba_{\mathrm{R,h}}(\varphi_{m}), 
\end{align}
where $\theta_{\mathrm{RB},m}^{\mathrm{Azi}}$ and $\theta_{\mathrm{RB},m}^{\mathrm{Ele}}$ are the azimuth and elevation angles of departure (AoD) of the $m$-th path, respectively. The RIS steering vectors along the vertical and horizontal directions are respectively given by
\begin{align}
	&\ba_{\mathrm{R,v}}(\psi_{m}) =\frac{1}{\sqrt{L_{\mathrm{v}}}}\left[1,e^{j \psi_{m} },\cdots, e^{j(L_{\mathrm{v}}-1)\psi_{m}} \right]^{\mathrm{T}} \nonumber
	\\& \ba_{\mathrm{R,h}}(\varphi_{m}) = \frac{1}{\sqrt{L_{\mathrm{h}}}}\left[1,e^{j \varphi_{m} },\cdots, e^{j(L_{\mathrm{h}}-1)\varphi_{m}} \right]^{\mathrm{T}},
\end{align}
where $\psi_{m}=2\pi d_{\mathrm{R,v}} \sin(\theta_{\mathrm{RB},m}^{\mathrm{Ele}})/\lambda_{\mathrm{c}}$ is the spatial frequency along the vertical direction with vertical spacing $d_{\mathrm{R,v}}$, and $\varphi_{m}=2\pi d_{\mathrm{R,h}}$ $\cos(\theta_{\mathrm{RB},m}^{\mathrm{Ele}}) \sin(\theta_{\mathrm{RB},m}^{\mathrm{Azi}})/ \lambda_{\mathrm{c}}$ is the horizontal spatial frequency with horizontal spacing $d_{\mathrm{R,h}}$. 
For simplicity, we reformulate $\bG$ in (\ref{G}) as
\begin{align} \label{G_simple}
	\bG = \bA_{\mathrm{B,RB}} \diag(\boldsymbol{\alpha}_{\mathrm{RB}}) \bA_{\mathrm{R,RB}}^{\mathrm{H}},
\end{align}
where $\bA_{\mathrm{B,RB}}=[\ba_{\mathrm{B}}(\phi_{1}),\cdots,\ba_{\mathrm{B}}(\phi_{M_{\mathrm{RB}}})] \in \mathbb{C}^{N \times M_{\mathrm{RB}}}$, $\bA_{\mathrm{R,RB}}=[\ba_{\mathrm{R}}(\theta_{\mathrm{RB},1}^{\mathrm{Azi}}, \theta_{\mathrm{RB},1}^{\mathrm{Ele}}),\cdots,\ba_{\mathrm{R}}(\theta_{\mathrm{RB},M_{\mathrm{RB}}}^{\mathrm{Azi}}, \theta_{\mathrm{RB},M_{\mathrm{RB}}}^{\mathrm{Ele}})] \in \mathbb{C}^{L \times M_{\mathrm{RB}}}$, and $\boldsymbol{\alpha}_{\mathrm{RB}}=\sqrt{\frac{NL}{M_{\mathrm{RB}}}}[\alpha_{\mathrm{RB},1},\cdots,\alpha_{\mathrm{RB},M_{\mathrm{RB}}}]^{\mathrm{T}} \in \mathbb{C}^{M_{\mathrm{RB}}\times 1}$.

Similarly, the channel $\mathbf{f}_k$ from the $k$-th UE to the RIS is expressed~as
\begin{align} \label{f_k}
	\mathbf{f}_k &= \sqrt{\frac{L}{M_{\mathrm{UR},k}}} \sum_{m=1}^{M_{\mathrm{UR},k}} \alpha_{\mathrm{UR},k,m} \ba_{\mathrm{R}}(\theta_{\mathrm{UR},k,m}^{\mathrm{Azi}}, \theta_{\mathrm{UR},k,m}^{\mathrm{Ele}}),
\end{align}
where $M_{\mathrm{UR},k}$ is the number of propagation paths for this channel, $\alpha_{\mathrm{UR},k,m} \sim \cC\cN(0, \sigma_{\mathrm{UR},k}^2)$ is the i.i.d. complex gain of the $m$-th path with variance $\sigma_{\mathrm{UR},k}^2$, and the RIS steering vector $\ba_{\mathrm{R}}(\theta_{\mathrm{UR},k,m}^{\mathrm{Azi}}, \theta_{\mathrm{UR},k,m}^{\mathrm{Ele}})$ is defined similarly to~(\ref{a_R}) with azimuth and elevation AoAs $\theta_{\mathrm{UR},k,m}^{\mathrm{Azi}}$ and $\theta_{\mathrm{UR},k,m}^{\mathrm{Ele}}$,~respectively. Based on (\ref{f_k}), $\mathbf{f}_k$ can be reformulated as 
\begin{align} \label{f_k_simple}
	\mathbf{f}_k = \bA_{\mathrm{R,UR},k} \boldsymbol{\alpha}_{\mathrm{UR},k},
\end{align}
where $\bA_{\mathrm{R,UR},k}=[\ba_{\mathrm{R}}(\theta_{\mathrm{UR},k,1}^{\mathrm{Azi}}, \theta_{\mathrm{UR},k,1}^{\mathrm{Ele}}),\cdots,\ba_{\mathrm{R}}(\theta_{\mathrm{UR},k,M_{\mathrm{UR},k}}^{\mathrm{Azi}},$ $\theta_{\mathrm{UR},k,M_{\mathrm{UR},k}}^{\mathrm{Ele}})]$ $\in \mathbb{C}^{L \times M_{\mathrm{UR},k}}$, and $\boldsymbol{\alpha}_{\mathrm{UR},k}=\sqrt{\frac{L}{M_{\mathrm{UR},k}}}[\alpha_{\mathrm{UR},k,1},$ $\cdots,\alpha_{\mathrm{UR},k,M_{\mathrm{UR},k}}]^{\mathrm{T}} \in \mathbb{C}^{M_{\mathrm{UR},k}\times 1}$.

\begin{figure}
	\centering
	\includegraphics[width=1.0\columnwidth]{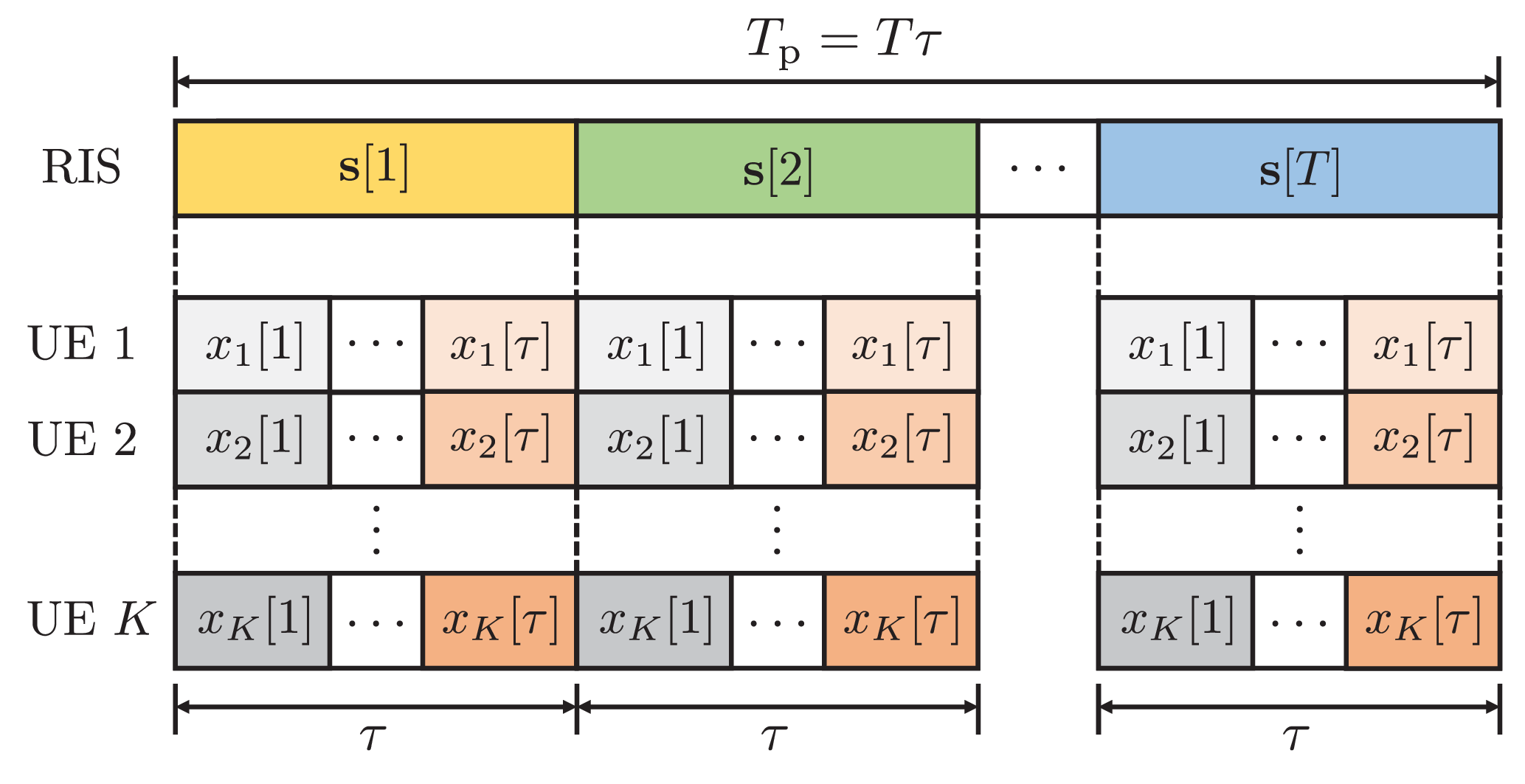}
	\caption{The uplink time-domain transmission design for estimating the cascaded channel.} \label{Pilot_pattern}	
\end{figure}

\section{Problem Formulation} \label{sec3}
In the following, we will consider two types of channel estimation problems for the above RIS-aided system: \textit{1) Cascaded channel estimation} and \textit{2) Individual RIS channel estimation.} The first problem is encountered in scenarios where all RIS elements are purely passive, and no information is available at the RIS from the training signals. Cascaded channel CSI is generally sufficient in many situations such as BS transceiver design, but there are applications where knowledge of the individual RIS channels is necessary \cite{Guo:2020, Zhang:2020, Hua:2023}. One approach for separately estimating the BS-RIS and RIS-UE channels involves the use of semi-passive elements at the RIS that can receive and process the training data to eliminate ambiguities normally present in the cascaded channel.

\subsection{Cascaded channel estimation} \label{sec3_1}
In this case, we assume the RIS consists of purely passive elements, i.e., $L_{\mathrm{p}}=L$ and $\bOmega=\b0_L$.
The uplink signal design used to estimate the cascaded channel is depicted in Fig.~\ref{Pilot_pattern}, where a block of length $T_{\mathrm{p}}$ is composed of $T$ subblocks of $\tau$ time slots each. Within each subblock, the RIS configuration remains unchanged, and all UEs transmit the same pilot sequences in each of the $T$ subblocks.
The RIS passive reflection vector and the transmit signal from the $k$-th UE in the $u$-th time slot of the $t$-th subblock are then respectively given~by
\begin{align} \label{pilot_protocol}
	\bs[t, u] &= \bs[t], \enspace 1\leq u \leq \tau, \nonumber \\
	x_k[t, u] &= x_k[u], \enspace 1\leq t \leq T, \enspace \forall k=1,\cdots,K. 
\end{align}
Based on (\ref{y_t}) and (\ref{pilot_protocol}), the signal received at the BS in the $u$-th time slot of the $t$-th subblock, $\by[t,u]=\by[(t-1)\tau+u]$, is expressed as
\begin{align} \label{y_t_u}
	\by[t,u] &= \sum_{k=1}^K \bG\diag(\bs[t])\mathbf{f}_k x_k[u] + \bn_{\mathrm{B}}[t,u] \nonumber
	\\ & = \bG\diag(\bs[t])\bF\bx[u] +  \bn_{\mathrm{B}}[t,u].
\end{align}
The observation matrix at the BS is obtained by stacking (\ref{y_t_u}) over $\tau$ time slots for the $t$-th subblock:
\begin{align} \label{Y_t}
	\bY[t] &= [\by[t,1],\cdots,\by[t,\tau]] \nonumber
	\\ &= \bG \diag(\bs[t]) \bF \bX_{\mathrm{C}} + \bN_{\mathrm{B}}[t],
\end{align}
where $\bX_{\mathrm{C}}=[\bx[1],\cdots,$ $\bx[\tau]] \in \mathbb{C}^{K \times \tau}$, and $\bN_{\mathrm{B}}[t] = [\bn_{\mathrm{B}}[t,1],\cdots,\bn_{\mathrm{B}}[t,\tau]] \in \mathbb{C}^{N \times \tau}$. Using the identities $\mathrm{vec}(\bM_1 \bM_2 \bM_3)=(\bM_3^{\mathrm{T}}\otimes \bM_1) \mathrm{vec}(\bM_2)$ and $\mathrm{vec}(\bM_1 \diag(\mathbf{m}) \bM_2) = (\bM_2^{\mathrm{T}} \diamond \bM_1 ) \mathbf{m}$, $\bY[t]$ in (\ref{Y_t}) can be vectorized as
\begin{align} \label{vec_Y_t}
	\mathrm{vec}(\bY[t]) =(\bX_{\mathrm{C}}^{\mathrm{T}} \otimes \bI_N)(\bF^{\mathrm{T}}\diamond \bG)\bs[t] + \mathrm{vec}(\bN_{\mathrm{B}}[t]).
\end{align}
Finally, the complete set of BS observations over all $T$ subblocks in (\ref{vec_Y_t}) is collected in the matrix $\bY \in \mathbb{C}^{N\tau \times T}$:
\begin{align} \label{Y}
	\bY &= [\mathrm{vec}(\bY[1]) ,\cdots, \mathrm{vec}(\bY[T])] \nonumber
	\\&=(\bX_{\mathrm{C}}^{\mathrm{T}} \otimes \bI_N)(\bF^{\mathrm{T}}\diamond \bG)\bS + \bN_{\mathrm{B}},
\end{align}
where $\bS=[\bs[1],\cdots,\bs[T]] \in \mathbb{C}^{L \times T}$, and $\bN_{\mathrm{B}}=[\mathrm{vec}(\bN_{\mathrm{B}}[1]),\cdots,\mathrm{vec}(\bN_{\mathrm{B}}[T])] \in \mathbb{C}^{N\tau \times T}$. 

From (\ref{Y}), let $\bC = \bF^{\mathrm{T}}\diamond \bG$ be the cascaded channel between the BS and the UEs.
Using the identity $\mathrm{vec}(\bM_1 \bM_2 \bM_3)=(\bM_3^{\mathrm{T}}\otimes \bM_1) \mathrm{vec}(\bM_2)$, the received signals at the BS in (\ref{Y}) can be vectorized~as
\begin{align} \label{y_c}
	\mathrm{vec}(\bY) &= (\bS^{\mathrm{T}} \otimes \bX_{\mathrm{C}}^{\mathrm{T}} \otimes \bI_N) \mathrm{vec}(\bC) + \mathrm{vec}(\bN_{\mathrm{B}}) \nonumber
	\\ & = \bar{\bS} \bc +\bn_{\mathrm{B}} \nonumber \\&\triangleq \by,
\end{align}
where $\bar{\bS} =(\bS^{\mathrm{T}} \otimes \bX_{\mathrm{C}}^{\mathrm{T}} \otimes \bI_N)$, $\bc = \mathrm{vec}(\bC)$, and $\bn_{\mathrm{B}} = \mathrm{vec}(\bN_{\mathrm{B}})$.
Our objective is to design a channel estimator for $\bc$ from low-resolution quantized observations of $\by$ using a hybrid analog and digital architecture at the BS. 
Specifically, let $\bB_{\bc} \in \mathbb{C}^{G_{\bc} \times NT\tau}$ be an analog combining matrix that projects $\by$ onto a lower-dimensional space $\mathbb{C}^{G_{\bc} \times 1}$ for which $G_{\bc} \leq NT\tau$, and let $\bD_{\bc} \in \mathbb{C}^{NKL \times G_{\bc}}$ be a digital processing matrix that reconstructs $\bc$. Our approach presented in Section IV will employ task-based quantization with the goal of minimizing the following MSE distortion between the true and estimated channel:
\begin{align} \label{problem_c_original}
	\min_{\bB_{\bc}, \bD_{\bc}, \gamma_{\bc}} \mathbb{E} \left[ \Vert \bc - \hat{\bc} \Vert_2^2 \right],
\end{align}
where $\hat{\bc}$ is the estimate of $\bc$, and $\gamma_{\bc}$ is the support for the~ADC.

\subsection{Estimation of individual RIS channels} \label{sec3_2}
When an RIS consists entirely of passive elements, it is not possible to extract the individual CSI components of the BS-RIS and RIS-UE channels from the cascaded channels without additional information. This is due to the inherent ambiguity in the cascaded channel, since $\bG \diag(\mathbf{f}_k) = (\bG \bLambda) (\bLambda^{-1}\diag(\mathbf{f}_k))$ for any invertible diagonal matrix $\bLambda$ \cite{Pan:2022}. To address this ambiguity, we exploit the availability of a few semi-passive elements in the RIS.
Here, our goal is to estimate \{$\bG$, $\bF$\} from low-resolution quantized observations of \{$\bY$, $\bZ$\}, employing a task-based approach that minimizes the MSE channel estimation error for both $\bG$ and $\bF$. 

Since the signals forwarded from the RIS to the BS are already quantized and only bear information about the RIS-UE link, jointly estimating both the RIS-UE and BS-RIS channels is a challenging task. For this reasons, we formulate a two-stage channel estimation approach, where in Stage I the RIS-UE channel $\bF$ is estimated based on the quantized observations from the semi-passive RIS elements, and in Stage II the BS-RIS channel $\bG$ is determined using the quantized observations at the BS and the estimate of $\bF$ obtained in Stage I. Since the signals received by the semi-passive elements are independent of the RIS phase shifts, we will assume $\tau=1$ in this case such that $T_{\mathrm{p}}=T$, and each UE transmits a different pilot symbol in each subblock.

\textit{1) Stage I:} In this stage, our goal is to design an estimator for $\bF$ based on the quantized observations at the semi-passive elements $\boldsymbol{\pi}_{\bz}$. Note that, consistent with the low cost and low power design of the RIS, we assume it has no analog combining capability. Consequently, the optimization problem involves designing only a digital processing matrix at the BS $\bD_{\mathbf{f}} \in \mathbb{C}^{KL \times LT}$, as follows:
\begin{align} \label{MSE_f_original}
	\min_{\bD_{\mathbf{f}}} \mathbb{E}[\Vert \mathbf{f} - \hat{\mathbf{f}}\Vert_2^2],
\end{align}
where $\hat{\mathbf{f}}$ is the estimate of $\mathbf{f}=\mathrm{vec}(\bF)$.

\textit{2) Stage II:} In this stage, the goal is to estimate $\bG$ based on the quantized version of $\bY$ and the estimate $\hat{\bF}$ obtained in Stage I.
Stacking the received signal vectors at the BS over the $T$ subblocks, the overall measurement matrix in (\ref{Y}) can be equivalently written as
\begin{align} \label{Y_individual}
	\bY &= [\by[1],\cdots,\by[T]] \nonumber \\&= \bG(\bS \odot \bF \bX_{\mathrm{I}}) + \bN_{\mathrm{B}},
\end{align}
where $\bN_{\mathrm{B}}=[\bn_{\mathrm{B}}[1],\cdots,\bn_{\mathrm{B}}[T]] \in \mathbb{C}^{N \times T}$.
Denoting $\bar{\bX}_{\mathrm{I}}=(\bS \odot \bF \bX_{\mathrm{I}})$, the observations at the BS in (\ref{Y_individual}) can be vectorized~as
\begin{align} \label{y}
	\mathrm{vec}(\bY) &= \left(\bar{\bX}_{\mathrm{I}}^{\mathrm{T}} \otimes \bI_N \right) \mathrm{vec}({\bG}) + \mathrm{vec}(\bN_{\mathrm{B}}) \nonumber
	\\ &= \bW_{\by} \bg + \bn_{\mathrm{B}} \nonumber
	\\ &\triangleq \by,
\end{align}
where $\bW_{\by}=\left(\bar{\bX}_{\mathrm{I}}^{\mathrm{T}} \otimes \bI_N \right)$, $\bg = \mathrm{vec}({\bG})$, and $\bn_{\mathrm{B}}=\mathrm{vec}(\bN_{\mathrm{B}})$.
In this case, given $\by$ and $\hat{\mathbf{f}}$, we aim to jointly design an analog combining matrix $\bB_{\bg} \in \mathbb{C}^{G_{\bg} \times NT}$ for which $G_{\bg} \leq NT$, a digital processing matrix $\bD_{\bg} \in \mathbb{C}^{NL \times G_{\bg}}$, and the support for the ADC $\gamma_{\bg}$ to minimize the MSE distortion between $\bg$ and the estimate $\hat{\bg}$.
The optimization problem is formulated assuming $\mathbf{f}$ is known and is given by
\begin{align} \label{Problem_g_original}
	\min_{\bB_{\bg}, \bD_{\bg}, \gamma_{\bg}} \mathbb{E} \left[ \Vert \bg - \hat{\bg} \Vert_2^2 \right \vert \mathbf{f}].
\end{align}

\section{Task-Based Quantization} \label{sec4}
The goal of task-based quantization \cite{Shlezinger:2019, Shlezinger:2019_2, Shlezinger:2020, Salamatoan:2019, Bernardo:2023} is to jointly design the hybrid analog and digital combining matrices and the quantization to minimize the MSE distortion between the true and estimated channel.
We focus on a system operating with identical scalar ADCs, referred to as a hardware-limited task-based quantizer \cite{Shlezinger:2019}, since a corresponding vector quantizer, despite its superior performance, would be computationally infeasible in practice for a large array. In the following, we will first explain the basic idea of the hardware-limited task-based quantizer and establish a theoretical basis for the proposed channel estimation design.

\subsection{Hardware-limited task-based quantizer} \label{sec4_1}
\begin{figure*}[t!]
	\centering
	\includegraphics[width=1.50\columnwidth]{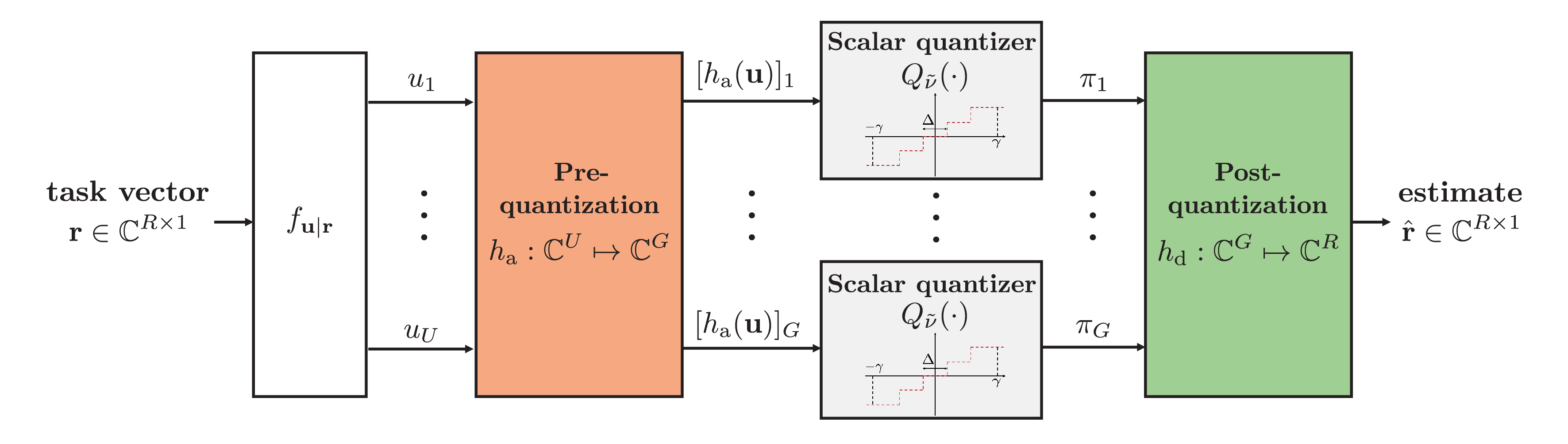}
	\caption{An example of a hardware-limited task-based quantization system.} \label{Hardware_limited_1} 
\end{figure*}
Fig. \ref{Hardware_limited_1} depicts a general hardware-limited task-based quantization system, where
$\br \in \mathbb{C}^{R \times 1}$ denotes the task vector we aim to recover, $\bu \in \mathbb{C}^{U \times 1}$ is the measurement vector, and the statistical relationship between $\br$ and $\bu$ is represented by the conditional probability $f_{\bu \vert \br}$. First, a pre-quantization mapping $h_{\mathrm{a}}(\cdot)$ carried out in the analog domain projects $\bu$ onto a lower-dimensional space $\mathbb{C}^{G \times 1}$ for which $G \leq U$. Subsequently, each component of $h_{\mathrm{a}}(\bu) \in \mathbb{C}^{G \times 1}$ is quantized element-wise using identical scalar ADCs.
Due to their useful statistical properties, the ADCs are modeled as non-subtractive uniform dithered quantizers \cite{Dithered_quantizer}, which facilitates a tractable analysis for the channel estimator design to be presented later.\footnote{Assuming a non-overloaded quantizer, the output of a uniform scalar ADC modeled by dithering can be equivalently described as the sum of the input and uncorrelated quantization noise. However, this characteristic can be approximately achieved in systems where dithering is not used for various types of inputs, such as Gaussian signals \cite{Widrow:1996}. The impact of dithering will be investigated in Section \ref{sec_numerical_cascaded}.}

We define $\tilde{\nu}=\lfloor \nu^{\frac{1}{2G}} \rfloor$, 
where $\log_2 \nu$ is the total number of quantization bits sampled by the system at any given time instant, and the individual ADC resolution is thus $\log_2 \tilde{\nu}$ bits. The ADC with dithering is modeled by the operation $Q_{\tilde{\nu}}(I_g) = q(\mathrm{Re}(I_g+\beta_g))+j\cdot q(\mathrm{Im}(I_g+\beta_g))$, $g=1,\cdots,G$.
Note that there are $2G$ ADCs to separately quantize the real and imaginary parts of $I_g$.
Here, $\{\beta_1,\cdots,\beta_G\}$ represent i.i.d. dither signals whose real and imaginary parts are uniformly distributed over $[-\Delta/2, \Delta/2]$ for quantization step size $\Delta = \frac{2 \gamma}{\tilde{\nu}}$ and mutually independent inputs, and $q(\cdot)$ is defined~as
\begin{align} \label{quantizer}
	q(I) = 
	\begin{cases}
		-\gamma + \Delta \left( \ell+\frac{1}{2} \right) \enspace & I -\ell \cdot \Delta + \gamma \in [0,\Delta], \\ & \ell \in \{0,\cdots,\tilde{\nu}-1\},
		\\ \mathrm{sign}(I)\left( \gamma - \frac{\Delta}{2} \right) \enspace &\vert I \vert > \gamma,
	\end{cases}
\end{align}
where $\gamma$ is the support for the ADC.
To guarantee that the inputs of the ADC in (\ref{quantizer}) fall within the range $[-\gamma, \gamma]$ with high probability, $\gamma$ is designed as some multiple $\eta$ of the maximum standard deviation of the inputs.
Finally, a post-quantization mapping $h_{\mathrm{d}}(\cdot)$ carried out in the digital domain reconstructs $\br$ based on the $G$ outputs of the identical scalar ADCs. Denoting the $g$-th output of the scalar ADC as $\pi_g = Q_{\tilde{\nu}}([h_{\mathrm{a}}(\bu)]_g)$, $\br$ is reconstructed as $\hat{\br} = h_{\mathrm{d}}(\boldsymbol{\pi}) \in \mathbb{C}^{R \times 1}$, where $\boldsymbol{\pi} = [\pi_1 ,\cdots, \pi_G]^{\mathrm{T}} \in \mathbb{C}^{G \times 1}$.

We focus on the problem of recovering a zero-mean random vector $\br$ from a zero-mean random measurement vector $\bu$, where all entries in both $\br$ and $\bu$ have finite variances. We use MSE distortion as the task recovery error, aiming to design $h_{\mathrm{a}}$, $h_{\mathrm{d}}$, and $\gamma$ to minimize the MSE between $\br$ and its estimate $\hat{\br}$. This leads to the following optimization problem:
\begin{align} \label{MSE_r}
	&\min_{h_{\mathrm{a}},h_{\mathrm{d},}, \gamma} \mathbb{E} \left[ \Vert \br - \hat{\br} \Vert_2^2 \right] \nonumber \\ &\mathop = \limits^{(a)} \mathbb{E}\left[\Vert \br - \tilde{\br} \Vert_2^2 \right] + \min_{h_{\mathrm{a}},h_{\mathrm{d}}, \gamma} \mathbb{E} \left[ \Vert \tilde{\br} - \hat{\br} \Vert_2^2 \right], 
\end{align}
where $\tilde{\br}=\mathbb{E}[\br \vert \bu]$ is the MMSE estimate of $\br$ based on $\bu$, and $(a)$ follows from the orthogonality principle stating that the MMSE estimate is uncorrelated with the estimation error \cite{Kay:book}.
The first term in (\ref{MSE_r}) is independent of the quantizer design, indicating the irreducible estimation error of $\br$ solely based on $\bu$. This implies that the optimization problem boils down to minimizing the second term, which represents the minimal distortion in quantizing the MMSE estimate of $\br$ \cite{Shlezinger:2019, Wolf:1970}, and this term would only be zero when the designed quantizer exactly recovers the MMSE estimate.
We will assume linear mappings for both $h_{\mathrm{a}}(\cdot)$ and $h_{\mathrm{d}}(\cdot)$, i.e., $h_{\mathrm{a}}(\bu) = \bB\bu$ with the analog combining matrix $\bB \in \mathbb{C}^{G \times U}$ and $h_{\mathrm{d}}(\boldsymbol{\pi}) = \bD \boldsymbol{\pi}$ with the digital processing matrix~$\bD \in \mathbb{C}^{R \times G}$. 

\subsection{Task-ignorant quantizer}
While task-based quantization systems are designed to minimize the MSE between $\br$ and $\hat{\br}$, the ADCs in task-ignorant systems are designed to simply reconstruct the given observation $\bu$ independently of the desired task, and the desired vector $\br$ is recovered from the quantized data by applying the MMSE estimator. 

\section{Cascaded Channel Estimation} \label{sec5}
In this section, we develop a cascaded channel estimator based on hardware-limited task-based quantization under finite bit-resolution constraints. 
Based on (\ref{MSE_r}), the optimization problem formulated in (\ref{problem_c_original}) boils down to
\begin{align} \label{Problem_c}
	\min_{\bB_{\bc}, \bD_{\bc}, \gamma_{\bc}} \mathbb{E} \left[ \Vert \tilde{\bc} - \hat{\bc} \Vert_2^2 \right],
\end{align}
where $\tilde{\bc}=\mathbb{E}[\bc \vert \by]$ is the MMSE estimate of $\bc$ given $\by$.

To solve (\ref{Problem_c}), analyzing the MMSE estimate $\tilde{\bc}$ is necessary. To facilitate the analysis, we provide the following lemma to reformulate~$\bc$.
\begin{lemma} \label{Lemma_c}
	The vectorized cascaded channel $\bc$ can be expressed as
	\begin{align} \label{c}
		\bc &= ((\bA_{\mathrm{R,UR}}^{\mathrm{T}} \diamond \bA_{\mathrm{R,RB}}^{\mathrm{H}})^{\mathrm{T}} \diamond \tilde{\bA}_{\mathrm{B,RB}}) (\boldsymbol{\alpha}_{\mathrm{UR}} \otimes \boldsymbol{\alpha}_{\mathrm{RB}}) \nonumber
		\\& \triangleq \bW_{\bc} \boldsymbol{\alpha}_{\bc},
	\end{align}
	where $\bA_{\mathrm{R,UR}} = [\bA_{\mathrm{R,UR},1},\cdots,\bA_{\mathrm{R,UR},K}]$, $\tilde{\bA}_{\mathrm{B,RB}} =\mathrm{blkdiag}(\boldsymbol{1}_{M_{\mathrm{UR},1}}^{\mathrm{T}},\cdots,\boldsymbol{1}_{M_{\mathrm{UR},K}}^{\mathrm{T}}) \otimes \bA_{\mathrm{B,RB}}$, $\boldsymbol{\alpha}_{\mathrm{UR}}^{\mathrm{T}} = [\boldsymbol{\alpha}_{\mathrm{UR},1}^{\mathrm{T}},\cdots,\boldsymbol{\alpha}_{\mathrm{UR},K}^{\mathrm{T}}]$, $\bW_{\bc}=(\bA_{\mathrm{R,UR}}^{\mathrm{T}} \diamond \bA_{\mathrm{R,RB}}^{\mathrm{H}})^{\mathrm{T}} \diamond \tilde{\bA}_{\mathrm{B,RB}}$, and $\boldsymbol{\alpha}_{\bc}=\boldsymbol{\alpha}_{\mathrm{UR}} \otimes \boldsymbol{\alpha}_{\mathrm{RB}}$.
	\begin{proof}
		See Appendix A.
	\end{proof}
\end{lemma}
Note that in general, the AoAs/AoDs in $\bW_{\bc}$ change much more slowly than the complex channel gains in $\boldsymbol{\alpha}_{\bc}$, and it can be assumed that $\bW_{\bc}$ remains fixed across multiple channel coherence blocks \cite{Geometric_1, Kim:2024}.
Based on this, we can assume that $\bW_{\bc}$ is deterministic, and thus the distribution of $\bc$ is based on a linear transformation of the random vector $\boldsymbol{\alpha}_{\bc}$ by $\bW_{\bc}$. Note that each element in $\boldsymbol{\alpha}_{\bc}$ is the product of independent complex channel gains that follow a zero-mean Gaussian distribution.
In \cite{Ware:2003}, it is shown that if $x_1 \sim \cC\cN(0, \sigma_1^2)$ and $x_2 \sim \cC\cN(0, \sigma_2^2)$, then the mean and variance of the product $y=x_1 x_2$ are zero and $\sigma_1^2 \sigma_2^2$, respectively. In our analysis below, we will approximate $y$ as a complex Gaussian random variable with this mean and variance.
Based on this result, $\boldsymbol{\alpha}_{\bc}$ approximately follows the following complex Gaussian distribution: $\boldsymbol{\alpha}_{\bc} \sim \cC\cN(\b0_{M_{\mathrm{RB}}M_{\mathrm{UR}}}, \bSigma_{\boldsymbol{\alpha}_{\bc}}$), where $M_{\mathrm{UR}}=\sum_{k=1}^K M_{\mathrm{UR},k}$, and $ \bSigma_{\boldsymbol{\alpha}_{\bc}} = \mathrm{blkdiag}(\bar{\sigma}_{\mathrm{RB}}^2 \bar{\sigma}_{\mathrm{UR},1}^2 \bI_{M_{\mathrm{RB}}M_{\mathrm{UR},1}},$ $\cdots,\bar{\sigma}_{\mathrm{RB}}^2 \bar{\sigma}_{\mathrm{UR},K}^2 \bI_{M_{\mathrm{RB}}M_{\mathrm{UR},K}})$ with $\bar{\sigma}_{\mathrm{RB}}^2 = \frac{NL}{M_{\mathrm{RB}}}\sigma_{\mathrm{RB}}^2$ and $\bar{\sigma}_{\mathrm{UR},k}^2=\frac{L}{M_{\mathrm{UR},k}}\sigma_{\mathrm{UR},k}^2$ denoting the variances of the elements in $\boldsymbol{\alpha}_{\mathrm{RB}}$ and $\boldsymbol{\alpha}_{\mathrm{UR},k}$, respectively.
This means that $\bc \sim \cC\cN(\b0_{NKL}, \bSigma_{\bc})$ approximately holds, where $\bSigma_{\bc}=\mathbb{E}[\bc \bc^{\mathrm{H}}]= \bW_{\bc} \bSigma_{\boldsymbol{\alpha}_{\bc}} \bW_{\bc}^{\mathrm{H}}$. Thus, the relationship between $\tilde{\bc}$ and $\by$ is linear, i.e., $\tilde{\bc}$ can be represented by $\tilde{\bc} = \bGamma_{\bc} \by$, where $\bGamma_{\bc}$ is given by
\begin{align} \label{Gamma_c}
	\bGamma_{\bc} &= \mathbb{E}\left[\bc\by^{\mathrm{H}}\right] \left(\mathbb{E}\left[\by\by^{\mathrm{H}}\right]\right)^{-1} \nonumber 
	\\&= \bSigma_{\bc} \bar{\bS}^{\mathrm{H}} \left(\bar{\bS} \bSigma_{\bc} \bar{\bS}^{\mathrm{H}} + \sigma_{\mathrm{B}}^2 \bI_{NT\tau}\right)^{-1}.
\end{align}

Based on the above discussions, we develop here a hardware-limited task-based quantization approach for estimating $\bc$ to minimize the MSE in (\ref{Problem_c}).
Denoting $\log_2 \nu_{\bc}$ as the total number of quantization bits used to estimate $\bc$, the resolution of the ADCs that separately quantize the real and imaginary parts of each element in $\bB_{\bc} \by$ is $\tilde{\nu}_{\bc}= \lfloor \nu_{\bc}^{\frac{1}{2G_{\bc}}} \rfloor$.
First, the digital processing matrix that minimizes the MSE in (\ref{Problem_c}) for a given analog combining matrix is found via the following lemma.
\begin{lemma} \label{Lemma_digital}
	For any analog combining matrix $\bB_{\bc}$, the digital processing matrix that minimizes the MSE in (\ref{Problem_c}) is
	\begin{align}
		\bD_{\bc}^{\mathrm{o}}(\bB_{\bc}) = \bGamma_{\bc} \bSigma_{\by} \bB_{\bc}^{\mathrm{H}} \left(\bB_{\bc} \bSigma_{\by} \bB_{\bc}^{\mathrm{H}} + \frac{4\gamma_{\bc}^2}{3\tilde{\nu}_{\bc}^2}\bI_{G_{\bc}} \right)^{-1},
	\end{align}
	and the resulting minimum MSE distortion is given~by
	\begin{align} \label{Lemma_digital_MSE}
		&\min_{\bD_{\bc}} \mathbb{E} \left[ \Vert \tilde{\bc} - \hat{\bc} \Vert_2^2 \right] = \trace\left(\bGamma_{\bc} \bSigma_{\by} \bGamma_{\bc}^{\mathrm{H}} \right) \nonumber \\& - \trace \left( \bGamma_{\bc} \bSigma_{\by} \bB_{\bc}^{\mathrm{H}} \left(\bB_{\bc} \bSigma_{\by} \bB_{\bc}^{\mathrm{H}} + \frac{4\gamma_{\bc}^2}{3\tilde{\nu}_{\bc}^2}\bI_{G_{\bc}} \right)^{-1} \bB_{\bc} \bSigma_{\by} \bGamma_{\bc}^{\mathrm{H}} \right),
	\end{align}
	where $\bSigma_{\by} = \mathbb{E}[\by \by^{\mathrm{H}}]$.
	\begin{proof}
		The proof is based on Lemma 1 in \cite{Shlezinger:2019}, and details are provided in Appendix B.
	\end{proof}
\end{lemma}

\begin{figure*}[t!] 
	\centering
	\includegraphics[width=1.50\columnwidth]{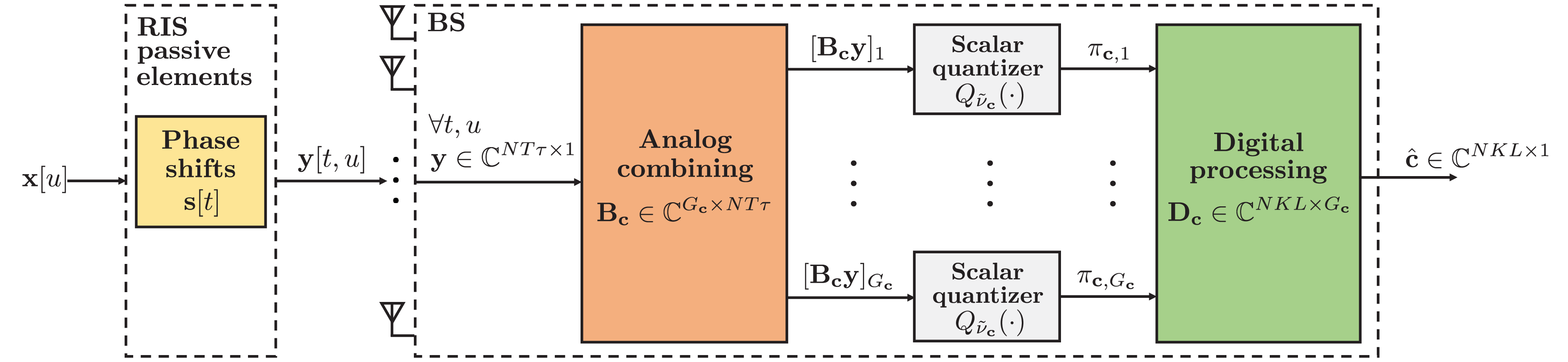}
	\caption{Architecture for cascaded channel estimation using hardware-limited task-based quantization.} \label{Hardware_limited_2} 
\end{figure*}

Using the results of Lemma \ref{Lemma_digital}, the following proposition characterizes the analog combining matrix that minimizes the MSE distortion in (\ref{Lemma_digital_MSE}).
\begin{proposition} \label{Proposition_analog}
	The optimal analog combining matrix that minimizes the MSE in (\ref{Lemma_digital_MSE}) is given by $\bB_{\bc}^{\mathrm{o}}=\bU_{\bc} \bLambda_{\bc} \bV_{\bc}^{\mathrm{H}} \bSigma_{\by}^{-\frac{1}{2}}$, where:
	\begin{enumerate}
		\item $\bV_{\bc} \in \mathbb{C}^{NT\tau \times NT\tau}$ is the matrix of right singular vectors of $\widetilde{\bGamma}_{\bc} \triangleq \bGamma_{\bc} \bSigma_{\by}^{\frac{1}{2}}$.
		\item $\bLambda_{\bc} \in \mathbb{C}^{G_{\bc} \times NT\tau}$ is a diagonal matrix with diagonal entries given by
		\begin{align} \label{Lambda_c}
			([\bLambda_{\bc}]_{g, g})^2 = \frac{4 \kappa_{\bc}}{3 \tilde{\nu}_{\bc}^2 \cdot G_{\bc}} \left( \zeta_{\bc} \cdot \lambda_{\widetilde{\bGamma}_{\bc},g} -1 \right)^+,
		\end{align}
		where $\kappa_{\bc} = \eta_{\bc}^2 \left( 1- \frac{2\eta_{\bc}^2}{3\tilde{\nu}_{\bc}^2} \right)^{-1}$, $\eta_{\bc}$ represents a design parameter to guarantee that the inputs to the ADCs fall within the desired dynamic range, $\{ \lambda_{\widetilde{\bGamma}_{\bc},g} \}$ are the singular values of $\widetilde{\bGamma}_{\bc}$ arranged in descending order, and $\zeta_{\bc}$ is chosen to satisfy
		\begin{align}
			\frac{4 \kappa_{\bc}}{3 \tilde{\nu}_{\bc}^2 \cdot G_{\bc}} \sum_{g=1}^{G_{\bc}} \left( \zeta_{\bc} \cdot \lambda_{\widetilde{\bGamma}_{\bc},g} -1 \right)^+ = 1.
		\end{align}
		\item $\bU_{\bc} \in \mathbb{C}^{G_{\bc} \times G_{\bc}}$ is a unitary matrix that forces $\bU_{\bc} \bLambda_{\bc} \bLambda_{\bc}^{\mathrm{H}} \bU_{\bc}^{\mathrm{H}}$ to have identical diagonal entries.
		\item The resulting support of the ADCs is $\gamma_{\bc} = \sqrt{\frac{\kappa_{\bc}}{G_{\bc}}}$.
	\end{enumerate}
	\begin{proof}
		The proof is based on Theorem 1 in \cite{Shlezinger:2019}, and details are provided in Appendix C.
	\end{proof}
\end{proposition}

The unitary matrix $\bU_{\bc}$ in Proposition \ref{Proposition_analog} can be computed using Algorithm 2.2 in \cite{Unitary_matrix}, which guarantees that each ADC input has an identical variance and allows for the same quantization rule to be applied to each channel.
By employing $\bLambda_{\bc}$, it is possible to balance the estimation and quantization errors by suppressing the less significant singular values $\{ \lambda_{\widetilde{\bGamma}_{\bc},g} \}$, following the water-filling type formula specified in (\ref{Lambda_c}).

Note that, as shown in Lemma \ref{Lemma_digital} and Proposition \ref{Proposition_analog}, the quantizer in the proposed channel estimator is jointly designed with both the analog combining matrix and the digital processing matrix to minimize the channel estimation error. In contrast, the performance metric for designing a standard task-ignorant quantizer is solely the accuracy of the digital representation with respect to its input. This implies that the proposed task-specific estimator design can achieve superior performance compared to task-ignorant designs, even under the same per-ADC bit resolution constraints.

The overall proposed cascaded channel estimation design is depicted in Fig. \ref{Hardware_limited_2}, where the $g$-th element of $\bB_{\bc} \by$ is quantized using the low-resolution ADC discussed in Section \ref{sec4_1}, and the output of the scalar ADC is denoted by $\pi_{\bc,g} \triangleq Q_{\tilde{\nu}_{\bc}}([\bB_{\bc} \by]_g)$. Subsequently, $\bc$ is reconstructed as $\hat{\bc} = \bD_{\bc} \boldsymbol{\pi}_{\bc}$, where $\boldsymbol{\pi}_{\bc}=[\pi_{\bc,1},\cdots,\pi_{\bc,G_{\bc}}]^{\mathrm{T}} \in \mathbb{C}^{G_{\bc} \times 1}$. Based on the estimate $\hat{\bc}$, the cascaded channel for the $k$-th UE, $\bC_k=\bG \diag(\mathbf{f}_k)$, is reconstructed as
\begin{align}
	\hat{\bC}_k = [\hat{\bC}]_{(k-1)N+1:kN,:} \; ,
\end{align}
where $\hat{\bC}=\mathrm{unvec}(\hat{\bc})$. Note that in order to minimize the MSE, the output dimension of the analog combining matrix $G_{\bc}$ should not be larger than the rank of the covariance matrix of $\tilde{\bc}$, according to Corollary 1 in~\cite{Shlezinger:2019}, and the optimal $G_{\bc}$ is equivalent to the number of non-zero singular values $\{ \lambda_{\widetilde{\bGamma}_{\bc},g} \}$.
This implies that, based on (\ref{Gamma_c}), the choice of $G_{\mathrm{c}}$ heavily depends on the rank of $\bSigma_{\bc}$, which is determined by the rank of $\bW_{\bc}$ in (\ref{c}) since $\bSigma_{\boldsymbol{\alpha}_{\bc}}$ is a full rank matrix. For further analysis, we provide the following lemma regarding the rank of $\bW_{\bc}$.
\begin{lemma} \label{Lemma_W_c}
	If the following conditions are satisfied, $\bW_{\bc}$ has full column rank $M_{\mathrm{RB}}M_{\mathrm{UR}}$:
	\begin{enumerate}
		\item $L \geq M_{\mathrm{RB}} M_{\mathrm{UR}}$,
		\item $\theta_{\mathrm{UR},k,i_k}^{\mathrm{Azi}} \neq \theta_{\mathrm{RB},j}^{\mathrm{Azi}}$, $\theta_{\mathrm{UR},k,i_k}^{\mathrm{Ele}} \neq \theta_{\mathrm{RB},j}^{\mathrm{Ele}}$, $\forall k \in \{1,\cdots K\}$, $\forall i_k \in \{1,\cdots,M_{\mathrm{UR},k} \}$, $\forall j= \{1,\cdots,M_{\mathrm{RB}} \}$.
	\end{enumerate}
	\begin{proof}
		See Appendix D.
	\end{proof}
\end{lemma}
In mmWave systems that employ an RIS, the RIS should have a large number of elements to overcome the multiplicative path-loss, and thus since the number of propagation paths for the RIS-related channels is limited, the first condition will generally be satisfied. Furthermore, the AoAs/AoDs of different propagation paths can be treated as independent continuous random variables, from which the second condition for Lemma \ref{Lemma_W_c} is satisfied with probability one \cite{Wang:2021, Hong:2022, Book:statistical}.

Based on the above discussion, in large-scale systems where $L$ is sufficiently large, the rank of the covariance matrix of $\tilde{\bc}$ depends on the rank of $\bW_{\bc}$ and is upper bounded by $M_{\mathrm{RB}}M_{\mathrm{UR}}$ for proper design of the UE pilot sequence $\bX_{\mathrm{C}}$ and the passive RIS reflection coefficients $\bS$, which defines $\bGamma_{\bc}$ in (\ref{Gamma_c}).
To investigate the number of training time slots required to achieve $\mathrm{rank}(\bGamma_{\bc})=M_{\mathrm{RB}}M_{\mathrm{UR}}$, we provide the following lemma.
\begin{lemma} \label{Lemma_training_overhead}
	If the conditions in Lemma \ref{Lemma_W_c} hold and both $\bS$ and $\bX_{\mathrm{C}}$ have their maximum ranks with $\tau \geq K$, the necessary conditions for $\mathrm{rank}(\bGamma_{\bc})=M_{\mathrm{RB}}M_{\mathrm{UR}}$ are given by
	\begin{enumerate}
		\item $NT\tau \geq M_{\mathrm{RB}}M_{\mathrm{UR}}$,
		\item $KT \geq M_{\mathrm{UR}}$.
	\end{enumerate}
	\begin{proof}
		See Appendix E.
	\end{proof}
\end{lemma}
We can see that the first condition in Lemma \ref{Lemma_training_overhead} is easily satisfied when the number of BS antennas is large. The second condition indicates it is sufficient that $T$ be at least the average number of propagation paths in the UE-RIS links, which is usually small in mmWave systems.
Consequently, unlike task-ignorant systems, the analog combining matrix derived from Proposition \ref{Proposition_analog} can reduce the dimensionality of the observation vector $\by$ at the BS from $NT\tau$ to $M_{\mathrm{RB}}M_{\mathrm{UR}}$ with a small training overhead, which significantly reduces quantization error when the total number of quantization bits at the BS is highly~limited.

In Proposition \ref{Proposition_analog}, the matrix $\bB_{\bc}$ linearly combines the elements of the vector $\by$ consisting of received signals at the BS over multiple time instances during the channel estimation phase.
In \cite{Hardware:1} it was shown that it is feasible to combine signals received over multiple time instances in the analog domain using the concept of virtual channel extension, in which an analog combining matrix for multiple time instances is constructed by sequentially reusing a simple hardware architecture that generates an analog combining matrix.
Furthermore, hardware prototypes for analog combining with dynamic weights have been demonstrated in \cite{Hardware:2, Hardware:3}. In \cite{Hardware:2}, dynamically adjustable complex gains were implemented using highly reconfigurable noise-canceling constant-Gm vector modulators, while in \cite{Hardware:3} analog vector multipliers were used to control the gain and phase of the analog signals.

Finally, we analyze the computational complexity required to implement the proposed cascaded channel estimator discussed thus far.
To simplify the analysis, we assume $T_{\mathrm{p}}\ll L$ and $ G_{\bc} = M_{\mathrm{RB}}M_{\mathrm{UR}} \ll NT_{\mathrm{p}}$.
To compute $\bGamma_{\bc}$, the dominant complexity comes from the matrix multiplications involved and is given by $\cO(N^3 K^2 L^2 T_{\mathrm{p}})$.
The complexity required to compute the analog combining matrix $\bB_{\bc}$ based on Proposition \ref{Proposition_analog} is $\cO(N^3 K L T_{\mathrm{p}}^2)$.
Computing the digital processing matrix $\bD_{\bc}$ in Lemma \ref{Lemma_digital} requires a complexity of $\cO(N^3 K L T_{\mathrm{p}}^2)$.
Note that, the remaining operations including matrix-vector multiplications to obtain $\hat{\bc}$ based on $\bB_{\bc}$ and $\bD_{\bc}$ have negligible computational complexity.
Thus, the total complexity of the proposed cascaded channel estimator is $\cO(N^3 K^2 L^2 T_{\mathrm{p}})$. 

\section{Individual RIS Channel Estimation} \label{sec6}
In this section, we design an estimator for the individual RIS-related channels using hardware-limited task-based quantization, where information from the semi-passive elements of the RIS is utilized, and low-resolution ADCs are present at both the BS and the RIS. As discussed in Section \ref{sec3_2}, we apply task-based quantization to the two-stage estimation problem, where in Stage I the RIS-UE channel $\bF$ is estimated based on the quantized observations from the semi-passive RIS elements, and in Stage II the BS-RIS channel $\bG$ is estimated using the quantized observations at the BS and the estimate of $\bF$ obtained in Stage I.

\subsection{Stage I: Estimation of $\bF$} \label{sec6_1}
We first develop a channel estimator for $\bF$ based on quantized observations at the semi-passive elements equipped with low-resolution ADCs.
To facilitate the analysis, we use a pseudo-measurement model as in \cite{In-soo:2023}, where a statistically equivalent expression for $\bZ$ in (\ref{Z_original}) is given by
\begin{align} \label{Z_psuedo}
	\hat{\bZ} &= [\hat{\bz}[1],\cdots,\hat{\bz}[T]] \nonumber \\ &= \bar{\bOmega} \odot \bF \bX_{\mathrm{I}} + \bN_{\mathrm{R}}.
\end{align}
In (\ref{Z_psuedo}), $\hat{\bZ}$ contains the same information about $\bF$ as $\bZ$ for the semi-passive elements, and no information about $\bF$ at the passive element locations. This implies that the quantized values of $\hat{\bZ}$ corresponding to the zeros of $\bar{\bOmega}$ can be chosen arbitrarily from among the possible ADC outputs. The vectorized representation of $\hat{\bZ}$ is given by
\begin{align}
	\mathrm{vec}(\hat{\bZ}) &= \diag(\mathrm{vec}(\bar{\bOmega}))(\bX_{\mathrm{I}}^{\mathrm{T}}\otimes \bI_L) \mathrm{vec}(\bF) + \mathrm{vec}(\bN_{\mathrm{R}}) \nonumber \\ &= \bW_{\hat{\bz}}\mathbf{f} + \bn_{\mathrm{R}} \nonumber \\ & \triangleq \hat{\bz},
\end{align}
where $\bW_{\hat{\bz}}=\diag(\mathrm{vec}(\bar{\bOmega}))(\bX_{\mathrm{I}}^{\mathrm{T}}\otimes \bI_L)$, $\mathbf{f}=\mathrm{vec}(\bF)$, and $\bn_{\mathrm{R}}=\mathrm{vec}(\bN_{\mathrm{R}})$.

Our goal is to design an estimator of $\mathbf{f}$ from $\hat{\bz}$ by minimizing the MSE distortion under finite bit-resolution constraints using hardware-limited task-based quantization. While in principle an analog combining matrix such as that in Proposition \ref{Proposition_analog} could be implemented at the RIS to obtain an accurate estimate of $\mathbf{f}$ as discussed in Section \ref{sec3_2}, we assume that no analog processing is performed at the RIS to maintain the low cost and power consumption of the RIS. Mathematically this is equivalent to setting the RIS analog combiner as $\bI_{LT}$.
In this case, the optimization problem boils down to solely designing a digital processing matrix $\bD_{\mathbf{f}}$ under finite bit-resolution constraints, leading to the following optimization problem based on the result in (\ref{MSE_r}):
\begin{align} \label{MSE_f}
	\min_{\bD_{\mathbf{f}}} \mathbb{E}[\Vert \tilde{\mathbf{f}} - \hat{\mathbf{f}}\Vert_2^2],
\end{align}
where $\tilde{\mathbf{f}}=\mathbb{E}[\mathbf{f}\vert \hat{\bz}]$ is the MMSE estimate of $\mathbf{f}$ given $\hat{\bz}$.
Note that the digital processing matrix $\bD_{\mathbf{f}}$ found in (\ref{MSE_f}) is applied at the BS after the RIS forwards $\boldsymbol{\pi}_{\bz}$ to the BS.

From (\ref{f_k_simple}), with fixed $\bA_{\mathrm{R,UR},k}$, $\mathbf{f}_k$ is a linear transformation of the random vector $\boldsymbol{\alpha}_{\mathrm{UR},k}$ by $\bA_{\mathrm{R,UR},k}$, and thus is distributed as $\mathbf{f} \sim \cC \cN(\b0_{KL}, \bSigma_{\mathbf{f}})$ with covariance matrix $\bSigma_{\mathbf{f}}=\mathrm{blkdiag}$ $(\bar{\sigma}_{\mathrm{UR,1}}^2 \bA_{\mathrm{R,UR},1}\bA_{\mathrm{R,UR},1}^{\mathrm{H}},\cdots,\bar{\sigma}_{\mathrm{UR},K}^2 \bA_{\mathrm{R,UR},K}$ $\bA_{\mathrm{R,UR},K}^{\mathrm{H}})$. This suggests that the MMSE estimate $\tilde{\mathbf{f}}$ is linear in $\hat{\bz}$, i.e., $\tilde{\mathbf{f}} =\bGamma_{\mathbf{f}}\hat{\bz}$, where 
\begin{align}
	\bGamma_{\mathbf{f}} &= \mathbb{E}[\mathbf{f} \hat{\bz}^{\mathrm{H}} ] (\mathbb{E}[\hat{\bz}\hat{\bz}^{\mathrm{H}}])^{-1} \nonumber \\
	&= \bSigma_{\mathbf{f}} \bW_{\hat{\bz}}^{\mathrm{H}} (\bW_{\hat{\bz}}\bSigma_{\mathbf{f}}\bW_{\hat{\bz}}^{\mathrm{H}}+\sigma_{\mathrm{R}}^2\bI_{LT})^{-1}.
\end{align}

Assume that each low-resolution ADC at the RIS has resolution $\tilde{\nu}_{\mathbf{f}}=\lfloor \nu_{\mathbf{f}}^{\frac{1}{2L_{\mathrm{a}}}} \rfloor$, where $\log_2 \nu_{\mathbf{f}}$ represents the total number of quantization bits available at the RIS, of which each RIS ADC produces $\log_2 \tilde{\nu}_{\mathbf{f}}$ bits.
The digital processing matrix minimizing the MSE distortion in (\ref{MSE_f}) for this case is characterized by the following corollary:
\begin{corollary} \label{Corollary_digital}
	Assuming no analog combining at the RIS, the digital processing matrix which minimizes the MSE in (\ref{MSE_f}) is given~by
	\begin{align}
		\bD_{\mathbf{f}}^{\mathrm{o}}(\bI_{LT}) = \bGamma_{\mathbf{f}} \bSigma_{\hat{\bz}} \left( \bSigma_{\hat{\bz}} + \frac{4\kappa_{\mathbf{f}}\sigma_{\hat{\bz},\mathrm{max}}^2}{3\tilde{\nu}_{\mathbf{f}}^2}\bI_{LT} \right)^{-1},
	\end{align}
	where $\kappa_{\mathbf{f}} = \eta_{\mathbf{f}}^2 \left( 1- \frac{2\eta_{\mathbf{f}}^2}{3\tilde{\nu}_{\mathbf{f}}^2} \right)^{-1}$ with $\eta_{\mathbf{f}}$ defined similarly to (\ref{Lambda_c}), $\bSigma_{\hat{\bz}}=\mathbb{E}[\hat{\bz}\hat{\bz}^{\mathrm{H}}]$, and
	$\sigma_{\hat{\bz},\mathrm{max}}^2 = \max_{i=1,\cdots,LT} [\bSigma_{\hat{\bz}}]_{i,i}$.
	\begin{proof}
		The proof follows directly from Lemma \ref{Lemma_digital}.
	\end{proof}
\end{corollary}

\begin{figure*}[t!]
	\centering
	\includegraphics[width=1.50\columnwidth]{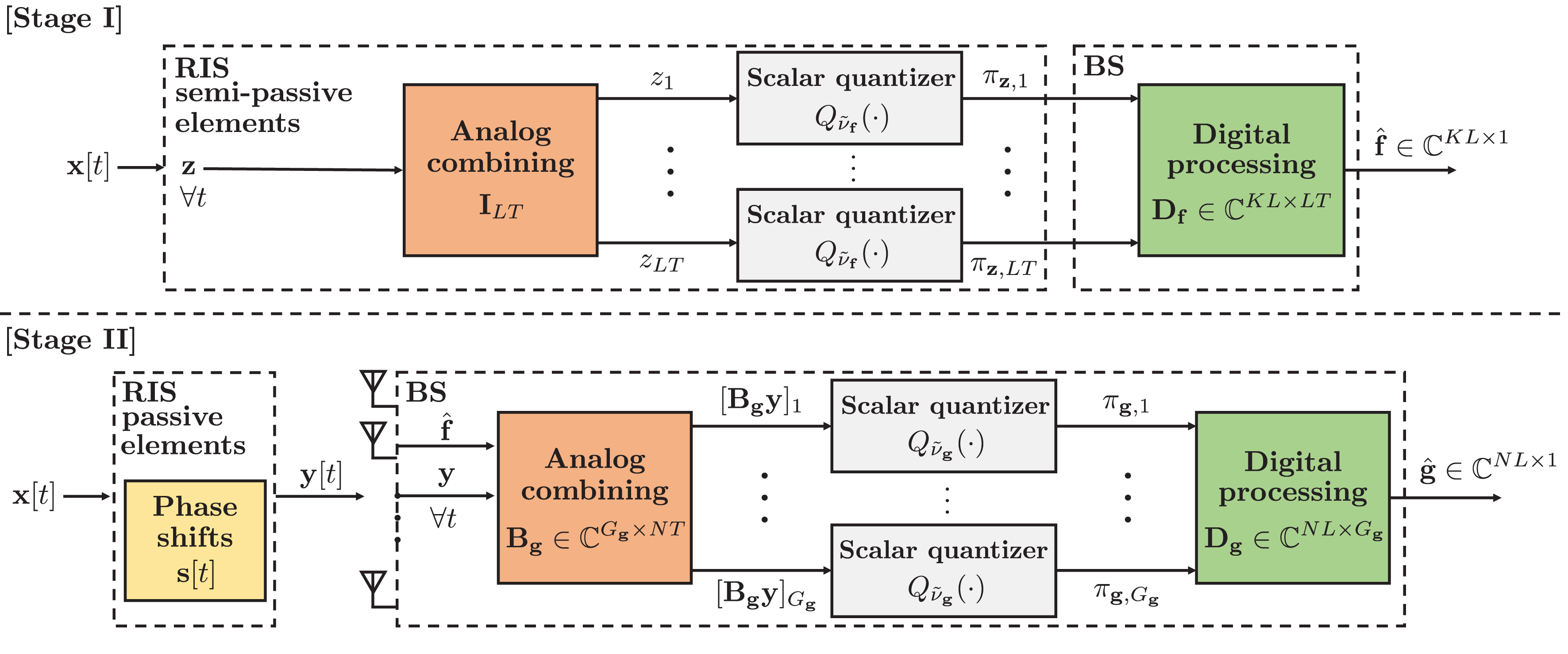}
	\caption{Architecture for estimating the individual RIS channels using hardware-limited task-based quantization.} \label{Hardware_limited_3} 
\end{figure*}

The overall estimation process for $\mathbf{f}$ is depicted in Fig. \ref{Hardware_limited_3}, where $\hat{\mathbf{f}}=\bD_{\mathbf{f}}\boldsymbol{\pi}_{\bz}$ and $\hat{\bF}=\mathrm{unvec}(\hat{\mathbf{f}})$.
Note that when the AoAs in the UE-RIS channels are distinct and $\bX_{\mathrm{I}}$ has its maximum rank, it can be shown that the necessary condition for $\mathrm{rank}(\bGamma_{\mathbf{f}})=M_{\mathrm{UR}}$ is $L_{\mathrm{a}}T \geq M_{\mathrm{UR}}$.

\subsection{Stage II: Estimation of $\bG$} \label{sec6_2}
In Stage II we estimate $\bG$ based on the quantized version of $\bY$ and the estimate $\hat{\bF}$ obtained in Stage I, where the optimization problem in (\ref{Problem_g_original}) is reformulated based on the result in~(\ref{MSE_r}):
\begin{align} \label{Problem_g}
	\min_{\bB_{\bg}, \bD_{\bg}, \gamma_{\bg}} \mathbb{E} \left[ \Vert \tilde{\bg} - \hat{\bg} \Vert_2^2 \right \vert \mathbf{f}],
\end{align}
where $\tilde{\bg}=\mathbb{E}[\bg \vert \by, \mathbf{f}]$ is the MMSE estimate of $\bg$ given $\by$ and $\mathbf{f}$.
From (\ref{G_simple}), $\bg$ can be represented as $\bg = (\bA_{\mathrm{R,RB}}^* \diamond \bA_{\mathrm{B,RB}})\boldsymbol{\alpha}_{\mathrm{RB}}$, implying that with fixed $\bA_{\mathrm{R,RB}}$ and $\bA_{\mathrm{B,RB}}$, $\bg$ is distributed as $\bg  \sim \cC \cN(\b0_{NL}, \bSigma_{\bg})$, with covariance $\bSigma_{\bg} = \bar{\sigma}
_{\mathrm{RB}}^2 (\bA_{\mathrm{R,RB}}^* \diamond \bA_{\mathrm{B,RB}}) (\bA_{\mathrm{R,RB}}^* \diamond \bA_{\mathrm{B,RB}})^{\mathrm{H}}$.
This implies that, given $\mathbf{f}$, $\tilde{\bg}$ is linear in $\by$ and can be represented as $\tilde{\bg}= \bGamma_{\bg \vert \mathbf{f}} \by$,~where
\begin{align} \label{Gamma_g_given_f}
	\bGamma_{\bg \vert \mathbf{f}} &= \mathbb{E}[\bg \by^{\mathrm{H}} \vert \mathbf{f} ] (\mathbb{E}[\by\by^{\mathrm{H}} \vert \mathbf{f}])^{-1} \nonumber \\
	&= \bSigma_{\bg} \bW_{\by}^{\mathrm{H}} (\bW_{\by}\bSigma_{\bg}\bW_{\by}^{\mathrm{H}}+\sigma_{\mathrm{B}}^2\bI_{NT})^{-1}.
\end{align}

Assuming a total of $\log_2 \nu_{\bg}$ quantization bits available at the BS to estimate $\bg$, the resolution of the individual ADCs is given by $\tilde{\nu}_{\bg}=\lfloor \nu_{\bg}^{\frac{1}{2G_{\bg}}} \rfloor$ bits.
Based on (\ref{Gamma_g_given_f}), the optimal $\bB_{\bg}$, $\bD_{\bg}$, and $\gamma_{\bg}$ that minimize the MSE in (\ref{Problem_g}) can be derived using the results from Proposition \ref{Proposition_analog} and Lemma \ref{Lemma_digital}.
Following Lemma \ref{Lemma_W_c}, it is straightforward to verify that $\bA_{\mathrm{R,RB}}^* \diamond \bA_{\mathrm{B,RB}}$ in $\bSigma_{\bg}$ has full column rank $M_{\mathrm{RB}}$ when $N \geq M_{\mathrm{RB}}$, implying that the analog combining matrix $\bB_{\bg}$ can reduce the dimensionality of the observations at the BS from $NT$ to $M_{\mathrm{RB}}$, given proper designs of $\bX_{\mathrm{I}}$ and $\bS$.
Furthermore, using an approach similar to that in Lemma \ref{Lemma_training_overhead}, it can be shown that a necessary condition for $\mathrm{rank}(\bGamma_{\bg \vert \mathbf{f}})=M_{\mathrm{RB}}$ is $NT \geq M_{\mathrm{RB}}$, and this condition can be satisfied even with $T=1$ when $N$ is large. This implies that, in large-scale systems, the RIS-BS channel can be accurately estimated with extremely small training overhead, while simultaneously achieving a significant reduction in the total number of quantization bits at the BS compared to task-ignorant estimators.

In large-scale systems, where a large number of BS antennas and RIS elements are deployed, the number of scalar ADCs required at the BS for the proposed algorithm to estimate the individual channels is $M_{\mathrm{UR}}$ times less than the number required by the proposed cascaded channel estimator in Section \ref{sec5}. This is a significant reduction, illustrating the clear advantage of the proposed individual channel estimator when a few semi-passive elements are available at the RIS.

The architecture for estimating $\bg$ is depicted in Fig. \ref{Hardware_limited_3}, where $\bGamma_{\bg \vert \mathbf{f}}$ in (\ref{Gamma_g_given_f}) is constructed using $\hat{\bF}$ obtained from Stage I. The estimate of $\bg$ is given by $\hat{\bg}=\bD_{\bg} \boldsymbol{\pi}_{\bg}$, where the $g$-th element in $\boldsymbol{\pi}_{\bg}=[\pi_{\bg,1},\cdots,\pi_{\bg,G_{\bg}}]^{\mathrm{T}} \in \mathbb{C}^{G_{\bg} \times 1}$ is $\pi_{\bg,g} \triangleq Q_{\tilde{\nu}_{\bg}}([\bB_{\bg}\by]_g)$, and $\bG$ can be reconstructed as $\hat{\bG}=\mathrm{unvec}(\hat{\bg})$.

The discussion above has assumed that the BS knows the AoAs/AoDs for the RIS-related channels since, as discussed in Section \ref{sec5}, these angles change relatively slowly. If the angles are unknown, they can be estimated given data from the semi-passive elements at the RIS using various techniques such as compressed sensing (CS) \cite{Taha:2021} or the method in \cite{Zhang:2021}. In Section~\ref{sec_numerical_practical}, we will compare the performance for cases where the angles must be estimated.

To analyze the computational complexity of the proposed individual channel estimator, we make assumptions similar to those used in the analysis of the proposed cascaded channel estimator, namely that $T \ll L$ and $G_{\bg} \ll NT$.
In Stage I, the complexity required to compute $\bGamma_{\mathbf{f}}$ is $\cO(KL^3T^2+K^2L^3T+L^3T^3)$, while computing the digital processing matrix $\bD_{\mathbf{f}}$ based on Corollary~\ref{Corollary_digital} requires a complexity of $\cO(KL^3T^2+L^3T^3)$, implying that the total complexity of Stage I is $\cO(KL^3T^2+K^2L^3T+L^3T^3)$.
In Stage II, computing $\bGamma_{\bg \vert \mathbf{f}}$ requires a complexity of $\cO(N^3 L^2 T)$.
Computing the analog combining and digital processing matrices has complexity $\cO(N^3 L T^2)$ and $\cO(N^3 L T^2)$, respectively, which leads to a total complexity for Stage II of $O(N^3 L^2 T)$.
Thus, the total computational complexity of the proposed individual channel estimator is $\cO(KL^3T^2+K^2L^3T+L^3T^3 + N^3 L^2 T)$.

\section{Numerical Results} \label{sec_numerical}
In this section, we investigate the performance of the proposed channel estimators based on hardware-limited task-based quantization. We assume a system with uplink carrier frequency $f_{\mathrm{c}}=24$ GHz, $N=16$ antennas at the BS, $L=100$ elements at the RIS with $L_{\mathrm{h}}=L_{\mathrm{v}}=10$, and $K=3$ UEs transmitting uncorrelated Gaussian signals.
The BS and RIS are located at (0 m, 0 m) and (20 m, 10 m), respectively, and the UEs are distributed around a circle centered at (40 m, 0 m) with radius 5 m.
The following system parameters are set based on the Dense Urban-eMBB scenario specified in ITU-R M.2412-0 \cite{ITU_R_M_2412_0}.
The UE transmit power is $P_k=\mbox{23 dBm}, \forall k=1,\cdots,K$. 
Assuming a noise spectral density $N_0=\mbox{-174 dBm/Hz}$, bandwidth $W=\mbox{80 MHz}$, and noise figure $\mathrm{NF}=\mbox{7 dB}$, the noise variances at the BS and RIS are $\sigma_{\mathrm{B}}^2 = \sigma_{\mathrm{R}}^2 = W \times N_0 \times \mathrm{NF}$.
Considering the line-of-sight dominant environments in mmWave bands, we set the the path-loss model as $\mathrm{PL}=31.4+20\log_{10}(r)$ dB, where $r$ represents the distance of the link in meters \cite{In-soo:2023}.
The antenna spacing at the BS and RIS is $d_{\mathrm{B}}=d_{\mathrm{R},\mathrm{v}}=d_{\mathrm{R},\mathrm{h}}=\frac{\lambda_{\mathrm{c}}}{2}$, where 
$\lambda_{\mathrm{c}}$ is the wavelength corresponding to $f_{\mathrm{c}}$.
For cascaded channel estimation, the number of time slots for each subblock is set to $\tau=K$.
The output dimensions of the analog combining matrices for cascaded and individual channel estimation are set to $G_{\bc}=M_{\mathrm{RB}} M_{\mathrm{UR}}$ and $G_{\bg} = M_{\mathrm{RB}}$, respectively.
Each RIS reflection coefficient in $\bS$ is randomly selected from the binary set $\{-1, 1\}$ to account for the hardware limitations of the RIS.
Unless otherwise specified, the number of propagation paths in the channel model is $M_{\mathrm{RB}}=M_{\mathrm{UR},1}=\cdots=M_{\mathrm{UR},K}=4$ considering the limited number of clusters in mmWave bands \cite{Li:2014}, and the number of subblocks for pilot training is $T=5$.

\subsection{Cascaded channel estimation performance} \label{sec_numerical_cascaded}
\begin{figure}
	\centering
	\includegraphics[width=1.0\columnwidth]{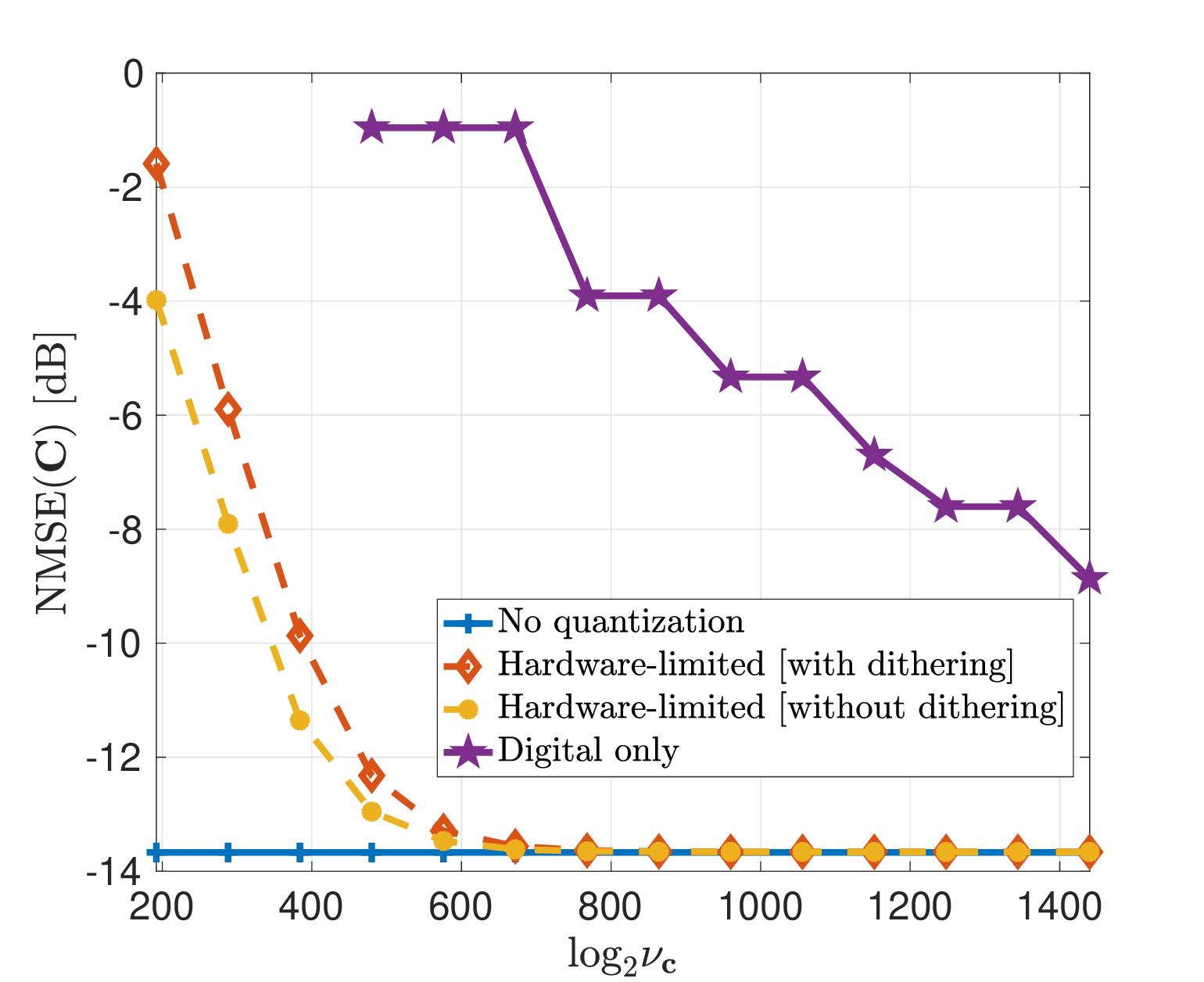}
	\caption{NMSE performance comparison for cascaded channel estimation versus total number of quantization bits.}	\label{NMSE_cascaded_bits}
\end{figure}

In this section, we evaluate the performance of the proposed cascaded channel estimator compared to the following baseline~schemes:
\begin{itemize}
	\item No quantization \cite{MMSE:book}: The MMSE estimate $\tilde{\bc}=\bGamma_{\bc} \by$ is applied without quantization. This estimate characterizes the minimum possible MSE distortion.
	\item Digital only: In this approach, the MMSE estimator $\bGamma_{\bc}$ is applied to the quantized observations without analog combining, i.e., the ADCs operate independently of the channel estimation task. In particular, the uniform quantizer proposed in \cite{Mo:2018,Max:1960} with the fixed resolution $\lfloor \nu_{\bc}^{\frac{1}{2NT\tau}} \rfloor$ is applied separately to the real and imaginary parts of each entry in $\by$.
\end{itemize}
We adopt the normalized MSE (NMSE) to measure the cascaded channel estimation error, which is defined as
\begin{align}
	\mathrm{NMSE}(\bC)= \mathbb{E}\left[ \frac{\Vert  \bC - \hat{\bC} \Vert_{\mathrm{F}} ^2}{\Vert \bC \Vert_{\mathrm{F}}^2} \right].
\end{align}

Fig. \ref{NMSE_cascaded_bits} illustrates the NMSE performance versus the total number of ADC quantization bits at the BS, $\log_2 \nu_{\bc}$.
It is observed that the NMSE difference between the proposed technique and the digital-only approach is quite large, suggesting that incorporating knowledge of the system task in the analog domain can yield significant performance improvement. This gain is mainly due to the design of the analog combining matrix that reduces the dimensionality of the input to the ADCs, allowing for more accurate quantization even with a small number of total quantization bits.
In particular, when each ADC uses at least six bits, the quantization error becomes negligible, and the NMSE of the proposed technique effectively approaches that of the MMSE estimate achievable with unlimited resolution ADCs.
For small $\log_2 \nu_{\bc}$, the proposed technique shows improved performance without dithering. As discussed in Section \ref{sec4_1}, Gaussian signals lead to approximately uncorrelated quantization noise without dithering \cite{Widrow:1996}. Since ADCs with dithering generally produce higher quantization noise than those without dithering, additional distortion is introduced to the inputs, resulting in a slight performance degradation for small $\log_2 \nu_{\bc}$, as discussed in \cite{Shlezinger:2019}. As $\log_2 \nu_{\bc}$ increases, this additional quantization noise becomes negligible, leading to nearly the same performance with or without dithering.

\begin{figure}
	\centering
	\includegraphics[width=1.0\columnwidth]{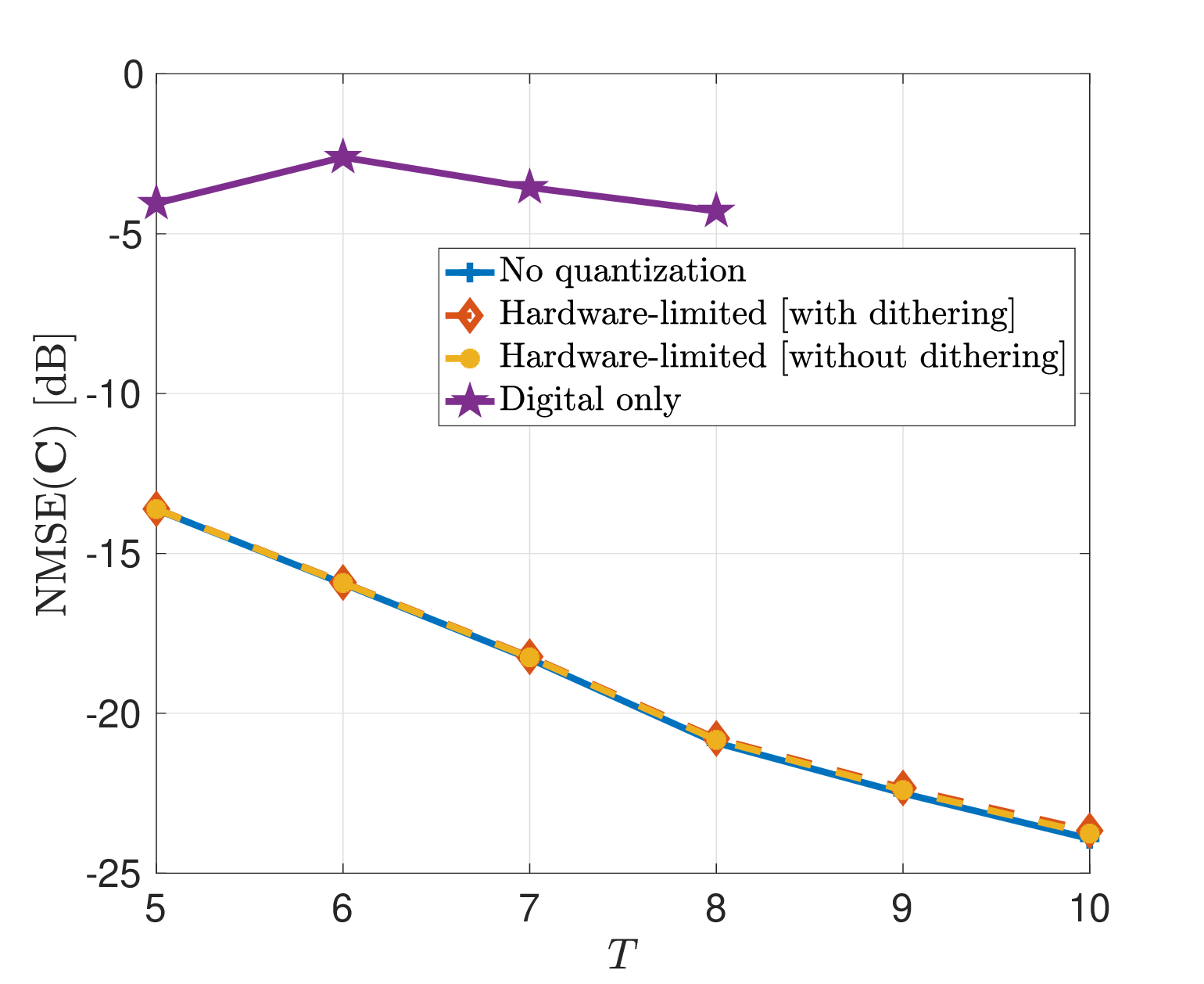}
	\caption{NMSE comparison for cascaded channel estimation versus number of pilot training subblocks.}
	\label{NMSE_cascaded_T}
\end{figure}
Fig. \ref{NMSE_cascaded_T} depicts the NMSE performance versus the number of subblocks $T$ assuming $\log_2 \nu_{\bc} = 800$.
The performance of the digital-only approach does not appreciably improve with $T$ since the ADC resolution per time instant decreases due to the increased dimension of the observation vector~$\by$.
Conversely, in the proposed technique, the resolution of each ADC is identical regardless of $T$ for a fixed $\log_2 \nu_{\bc}$, resulting in comparable performance to the ideal system without quantization when $\log_2 \nu_{\bc}$ is sufficiently large.  

\subsection{Individual RIS channel estimation performance}
Next we evaluate the performance achieved by the proposed two-stage approach for estimating the individual RIS channels.
In this case, we will use the following definitions of NMSE to quantify the estimation error for the individual channels:
\begin{align}
	&\mathrm{NMSE}(\bF)= \mathbb{E}\left[ \frac{\Vert  \bF - \hat{\bF} \Vert_{\mathrm{F}} ^2}{\Vert \bF \Vert_{\mathrm{F}}^2} \right],
	\\& \mathrm{NMSE}(\bG)= \mathbb{E}\left[ \frac{\Vert  \bG - \hat{\bG} \Vert_{\mathrm{F}} ^2}{\Vert \bG \Vert_{\mathrm{F}}^2} \right].
\end{align}
As in the previous section, approaches without quantization are used as the baselines. 
Specifically, in Stage I the no-quantization approach utilizes the MMSE estimate $\tilde{\mathbf{f}}=\bGamma_{\mathbf{f}} \bz$ without quantization, while in Stage II the MMSE estimate $\tilde{\bg}=\bGamma_{\bg \vert \mathbf{f}} \by$ is employed without quantization based on perfect knowledge of $\bF$.

\begin{figure}
	\centering
	\includegraphics[width=1.0\columnwidth]{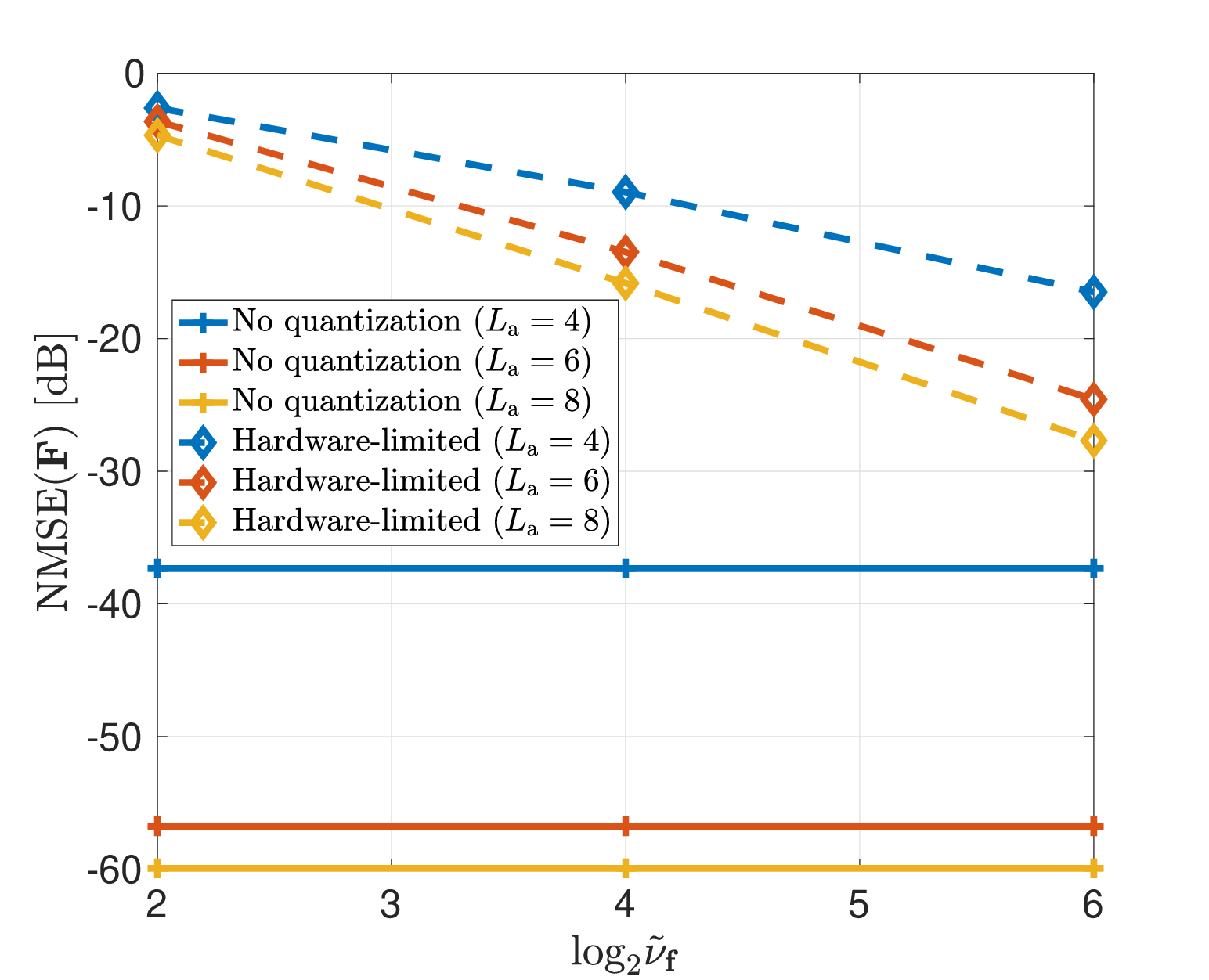}
	\caption{NMSE comparison of the RIS-UE link channel estimates versus the number of quantization bits at each semi-passive element.}	\label{NMSE_bits_F}
\end{figure}
In Fig. \ref{NMSE_bits_F}, we evaluate the NMSE performance for estimation of $\bF$ versus the ADC resolution $\log_2 \tilde{\nu}_{\mathbf{f}}$ of the semi-passive elements, whose locations are randomly chosen and fixed per iteration. The overall NMSE performance improves with the number of semi-passive elements $L_{\mathrm{a}}$ due to the increased number of observations available for estimation. Since no analog combining occurs at the RIS, the NMSE of the proposed technique is unable in these cases to achieve performance similar to the solution without quantization.
Nevertheless, when $L_{\mathrm{a}}=8$ and $\log_2 \tilde{\nu}_{\mathbf{f}}=6$, the NMSE of the proposed technique is approximately -30 dB, indicating that the channel is estimated with high accuracy. 

\begin{figure}
	\centering
	\includegraphics[width=1.0\columnwidth]{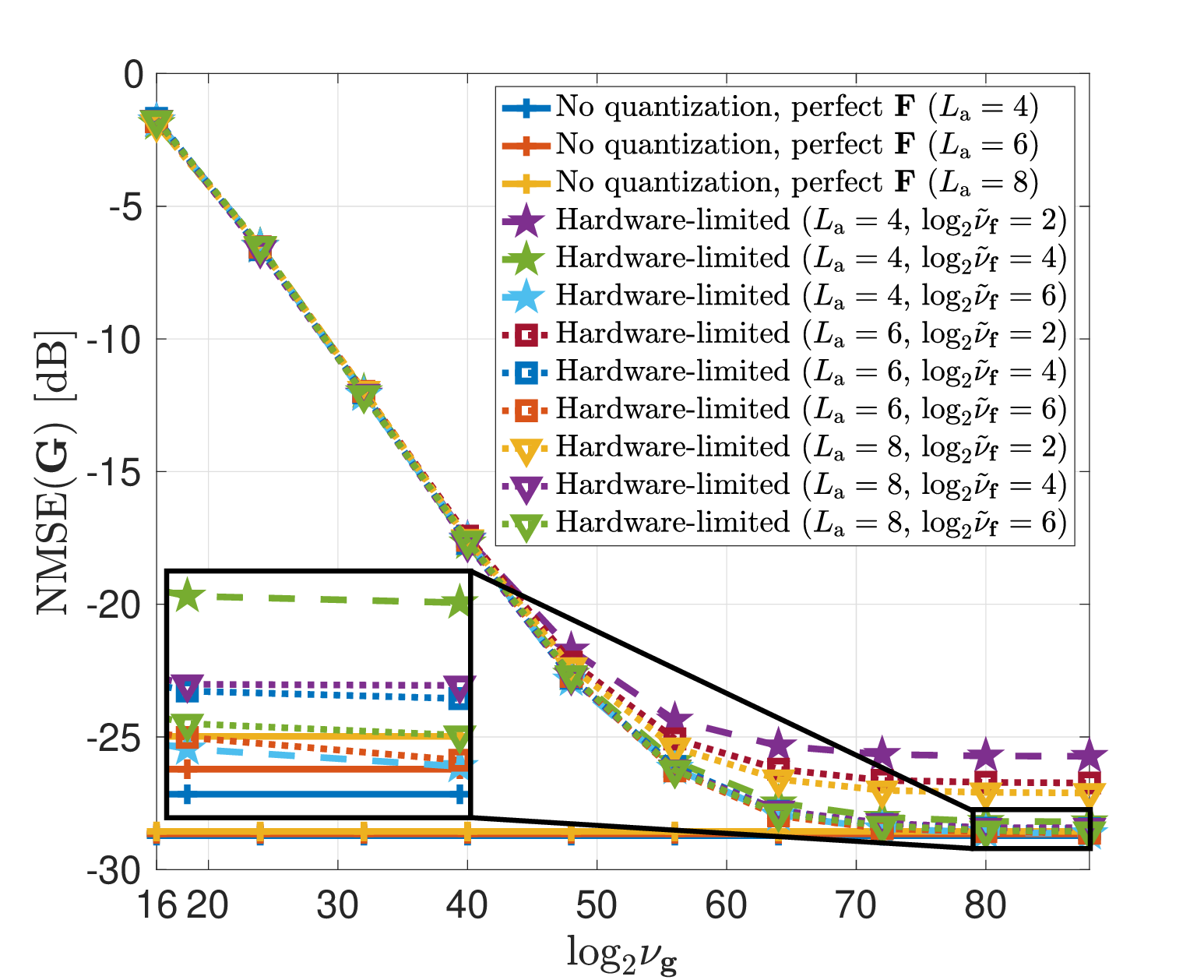}
	\caption{NMSE comparison of the BS-RIS channel estimate versus number of quantization bits at the BS.}	\label{NMSE_bits_G}
\end{figure}

In Fig. \ref{NMSE_bits_G}, we investigate the NMSE performance for estimation of $\bG$ versus the number of quantization bits $\log_2 \nu_{\bg}$ at the BS. When applying the proposed technique, the estimates of $\bF$ obtained from Stage I based on different $L_{\mathrm{a}}$ and $\log_2 \tilde{\nu}_{\mathbf{f}}$ are employed. We see that the NMSE without quantization increases slightly with $L_{\mathrm{a}}$ due to the reduced number of observations obtained through the passive elements.
Unlike Stage I for estimating $\bF$, the BS employs analog combining, resulting in a reduced NMSE difference between the performance of the proposed technique and that achievable without quantization when $\log_2 \nu_{\bg}$ is sufficiently large.
Despite its sub-optimal performance in estimating $\bF$ as seen in Fig. \ref{NMSE_bits_F}, the NMSE of the proposed technique for estimating $\bG$ effectively approaches the no-quantization lower bound based on perfect knowledge of $\bF$ when at least a 4-bit ADC is employed at each semi-passive RIS element, even for small $L_{\mathrm{a}}$.
This demonstrates that quality of the $\bF$ estimate obtained by the proposed technique is sufficient to match the performance of systems using unlimited resolution ADCs, even with only a small number of semi-passive elements.

\begin{figure}
	\centering
	\includegraphics[width=1.0\columnwidth]{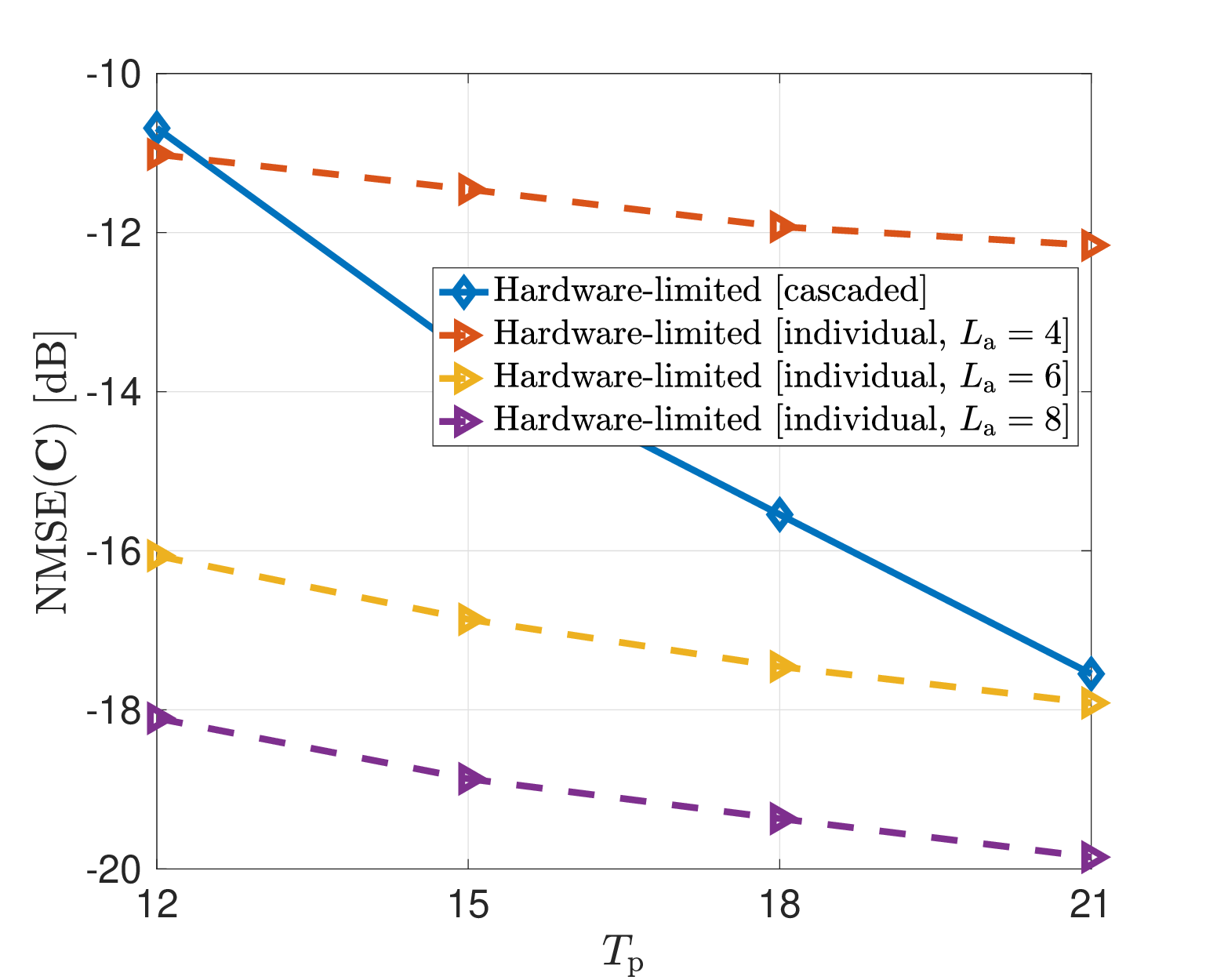}
	\caption{NMSE comparison for cascaded channel estimation versus number of time slots.}
	\label{NMSE_cascaded_vs_individual}
\end{figure}

We investigate the NMSE performance of the proposed cascaded and individual channel estimators versus the number of time slots $T_{\mathrm{p}}$ in Fig. \ref{NMSE_cascaded_vs_individual}. In this case, the BS requires a total of $\log_2 \nu_{\bc} = 576$ quantization bits for the cascaded channel estimator and $\log_2 \nu_{\bg}=48$ bits for the individual channel estimator, while each semi-passive element employs $\log_2 \tilde{\nu}_{\mathbf{f}}=4$ bits. For individually estimated RIS channels, the cascaded channel estimate is constructed from the individual  estimates. For small $L_{\mathrm{a}}$, the proposed cascaded channel estimate outperforms the one formed from the individual estimates since the limited number of observations at the semi-passive elements degrades the estimation performance. However, when at least $L_{\mathrm{a}}=6$ semi-passive elements are employed, the proposed individual channel estimator achieves performance superior to the cascaded channel estimate, with a significant reduction in the total number of quantization bits at the BS.

\subsection{Performance with estimated angles} \label{sec_numerical_practical}
The examples thus far have assumed that the BS has perfect knowledge of the AoAs/AoDs for the RIS-related channels since these angles vary relatively slowly compared to the complex channel gains, implying that they can be accurately estimated. Here we consider a scenario where these angles need to be estimated to demonstrate the robustness of the systems based on cascaded and individual channel estimation. We will consider the following baseline approaches:
\begin{itemize}
	\item LS: The well-known least squares (LS) estimator is applied to quantized observations at the BS to estimate the cascaded channel without the RIS semi-passive elements.
	\item DS-OMP \cite{Wei:2021}: This approach employs the double-structured OMP (DS-OMP) algorithm to estimate the cascaded channel leveraging the fact that the BS-RIS link channel is shared across the cascaded channels of all UEs. In this approach, the BS is equipped with infinite resolution ADCs without the RIS semi-passive elements.
	\item LS-OMP \cite{Han:2025}: This approach uses the LS-OMP algorithm for cascaded channel estimation with low-resolution ADCs at the BS without the RIS semi-passive elements, taking into account the potential leakage caused by grid mismatch.
	\item BiG-AMP \cite{Wang:2023}: This approach uses bilinear matrix completion to estimate the cascaded channel with low-resolution ADCs at the BS using known prior distributions, implemented with a modified BiG-AMP algorithm without the RIS semi-passive elements.
    \item ESPRIT-LS \cite{Zhang:2021}: This approach uses the signals received by the semi-passive elements to estimate the angles related to the BS-RIS and RIS-UE channels using the estimation of signal parameter via rotational invariance technique (ESPRIT). The complex channel gains are subsequently estimated using LS. We assume that
	the semi-passive elements are arranged as a $\sqrt{L_{\mathrm{a}}} \times \sqrt{L_{\mathrm{a}}}$ block at the bottom corner of the RIS.
	\item VI-SBL \cite{In-soo:2023}: This approach employs variational inference (VI) in the sparse Bayesian learning (SBL) framework to approximate the posterior distribution of the channel based on received signals at both the BS and the RIS semi-passive elements.
	\item GMMV-AMP \cite{Cao:2024}: This approach solves a CS-based GMMV problem together with matrix completion based on a system with RIS semi-passive elements equipped with low-resolution ADCs. It is implemented via a hierarchical message passing algorithm based on AMP-like approximations.
\end{itemize}
In the proposed technique, denoted as ESPRIT-hardware-limited, the AoAs/AoDs for all links are estimated using the ESPRIT-based approach, and subsequently the proposed hardware-limited task-based quantization algorithm is applied to reconstruct the channels using these~estimates. 
Note that in VI-SBL and GMMV-AMP, the BS is assumed to have unlimited resolution ADCs, and the locations of the semi-passive elements are randomly chosen at each time instance.

\begin{figure}
	\centering
	\includegraphics[width=1.0\columnwidth]{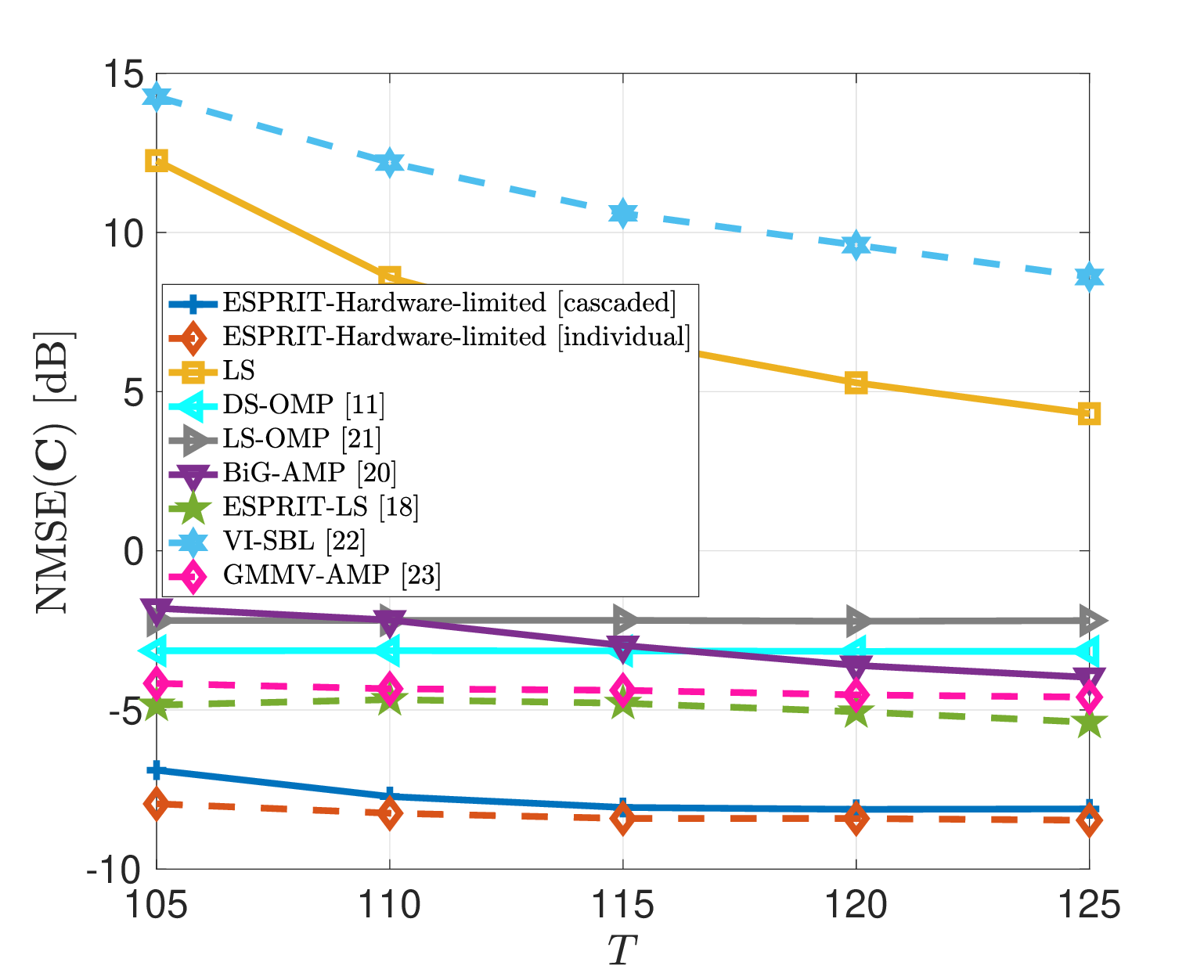}
	\caption{NMSE comparison of cascaded channel estimates versus number of time slots with estimated angles.}
	\label{NMSE_cascaded_estimated_angles}
\end{figure}

Fig. \ref{NMSE_cascaded_estimated_angles} plots the NMSE of the cascaded channel estimates versus the number of time slots  $T_{\mathrm{p}}=T$ assuming a single UE, $L_{\mathrm{a}}=9$, $M_{\mathrm{RB}}=1$, $M_{\mathrm{UR}}=2$, $\log_2 \nu_{\bc}=24$ bits, $\log_2 \nu_{\bg}=12$ bits, and $\log_2 \tilde{\nu}_{\mathbf{f}}=6$ bits.
For the LS, LS-OMP, and BiG-AMP approaches, 6-bit ADCs are employed at the BS. In the proposed channel estimators, the AoAs/AoDs are estimated during $T_{\mathrm{ESPRIT}}=100$ time slots, and task-based quantization is applied at the BS during the remaining time slots. While the estimates of the number of propagation paths for the RIS-related channels may differ from the ground truth, if the estimates match the ground truth, the number of quantization bits per ADC at the BS is six bits for the proposed channel estimators, the same as for the baseline algorithms.
From Fig. \ref{NMSE_cascaded_estimated_angles},  we see that the proposed channel estimators achieve the lowest NMSE of all the considered approaches including the CS-based algorithms with a small training overhead and a significant reduction in the total number of quantization bits at the BS, demonstrating the advantages of task-based quantization over conventional approaches.
Since the algorithm in \cite{Zhang:2021} uses the results of the eigenvalue decomposition of a sample covariance matrix, quantization effects resulting from using low-resolution ADCs degrade the accuracy of the angle estimates, making the performance of ESPRIT-based approaches, including the proposed channel estimators, relatively stable compared to other algorithms as $T$ increases.
However, the performance gap between the proposed channel estimators and ESPRIT-LS clearly shows the advantage of task-based quantization. In particular, the proposed estimators provide more accurate estimates of the complex path gains compared to ESPRIT-LS even for small $T$.
BiG-AMP, VI-SBL, and GMMV-AMP focus on approximating the posterior distributions of the channels rather than specific channel parameters such as the AoAs/AoDs. Thus these approaches will require a larger number of observations to accurately estimate the channels.
The performance of DS-OMP and LS-OMP is inferior to some of the baselines since the cascaded channel gains are not Gaussian distributed, and grid mismatch becomes more severe for these algorithms in large-scale systems.
Note that, as discussed in Sections \ref{sec5} and \ref{sec6}, although the proposed channel estimators have relatively high computational complexity, they require a very small total number of quantization bits, which highlights their clear advantages over task-ignorant~approaches.

\begin{figure}
	\centering
	\includegraphics[width=1.0\columnwidth]{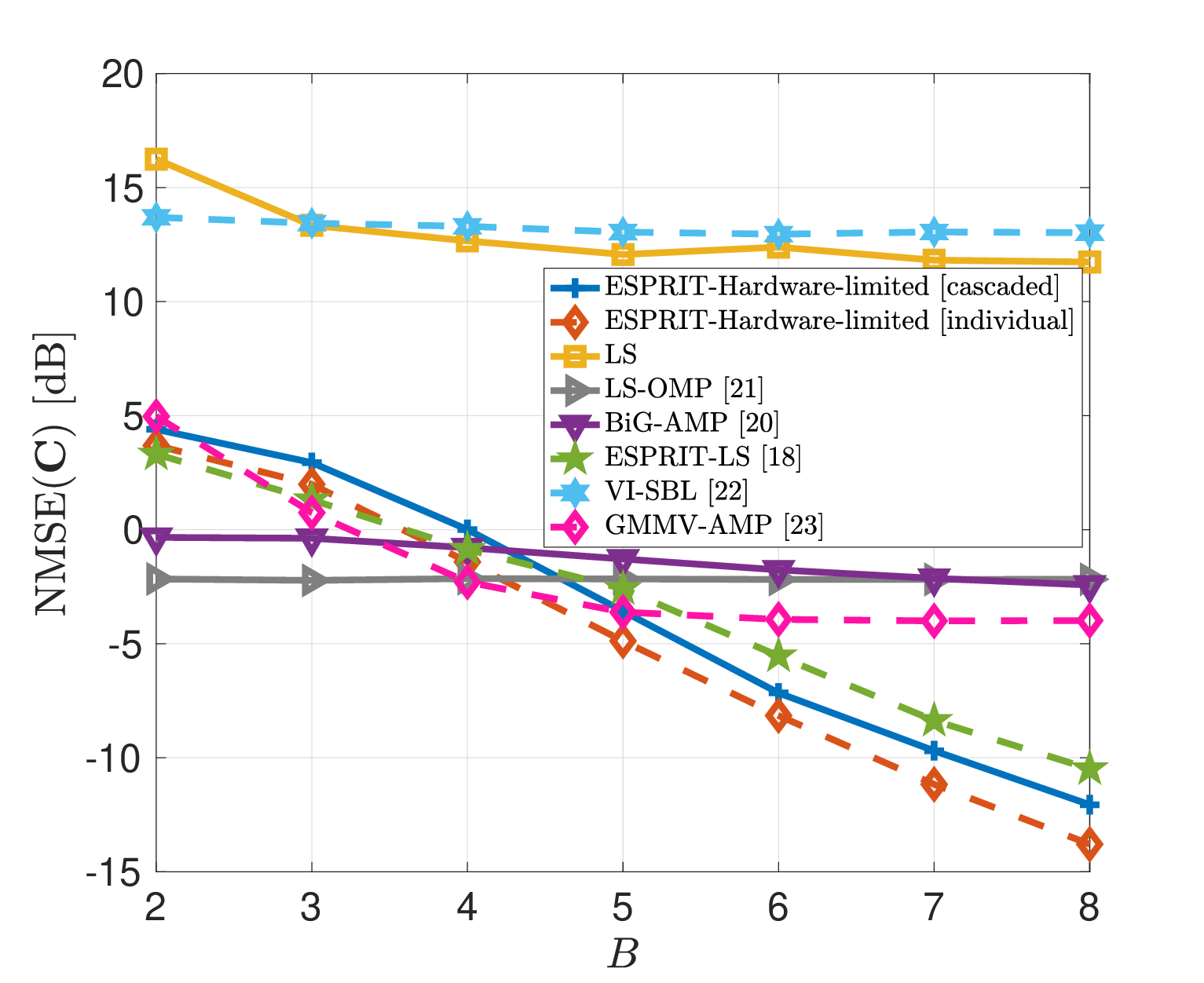}
	\caption{NMSE comparison for cascaded channel estimates versus per-ADC bit resolution with estimated angles.}
	\label{NMSE_cascaded_estimated_angles_per_ADC_bits}
\end{figure}
In Fig.~\ref{NMSE_cascaded_estimated_angles_per_ADC_bits}, we investigate algorithm performance versus the per-ADC bit resolution $B$ used to estimate the cascaded channel, where $L_{\mathrm{a}}=9$, $M_{\mathrm{RB}}=1$, $M_{\mathrm{UR}}=2$, $T=105$, and for the proposed channel estimators, $\log_2 \nu_{\bc}=2 M_{\mathrm{RB}} M_{\mathrm{UR}} B$ bits for the cascaded channel estimator and $\log_2 \nu_{\bg}=2 M_{\mathrm{RB}} B$ bits for the individual channel case. This implies that when the number of propagation paths is accurately estimated, the per-ADC bit resolution is the same for all approaches, except for VI-SBL and GMMV-AMP which use infinite-resolution ADCs at the BS.
We can see that when $B$ is at least five bits, the proposed channel estimators again show the lowest NMSE, achieved with significantly fewer total quantization bits at the~BS.

\section{Conclusion} \label{conclusion}
In this paper, we developed channel estimators for RIS-aided mmWave MU-SIMO systems using hardware-limited task-based quantization.
In the proposed approaches, an analog combining matrix, a digital processing matrix, and the support for the ADCs are jointly designed to minimize the MSE distortion of the channel estimate. In one approach, the cascaded channel is estimated based only on quantized observations at the BS. In the other, the observations at the BS are augmented by quantized observations at a few semi-passive elements at the RIS, and estimates of the individual RIS-related channels are obtained. Numerical results verify that, with relatively low-resolution ADCs, the proposed channel estimators can closely approach the performance of the MMSE estimate without quantization and outperform a purely digital approach. Furthermore, the proposed estimators are shown to outperform baseline approaches with a low training overhead in scenarios requiring angle estimates at the RIS.

\appendix{
	\subsection{Proof of Lemma \ref{Lemma_c}}
 	From (\ref{Y}), the cascaded channel is $\bC = \bF^{\mathrm{T}} \diamond \bG$, and $\bF$ can be reformulated as $\bF = [\mathbf{f}_1, \cdots, \mathbf{f}_K] = \bA_{\mathrm{R,UR}} \bLambda_{\mathrm{UR}}$, where $\bLambda_{\mathrm{UR}}=\mathrm{blkdiag}(\boldsymbol{\alpha}_{\mathrm{UR},1}, \cdots, \boldsymbol{\alpha}_{\mathrm{UR},K})$. Based on the expressions (\ref{G_simple}) and (\ref{f_k_simple}), $\bC$ can be rewritten as
	\begin{align} \label{cascade_reformulated}
		\bC &= (\bLambda_{\mathrm{UR}}^{\mathrm{T}}\bA_{\mathrm{R,UR}}^{\mathrm{T}}) \diamond (\bA_{\mathrm{B,RB}} \diag(\boldsymbol{\alpha}_{\mathrm{RB}}) \bA_{\mathrm{R,RB}}^{\mathrm{H}}) \nonumber
		\\ &\mathop = \limits^{(a)} (\bLambda_{\mathrm{UR}}^{\mathrm{T}}\otimes \bA_{\mathrm{B,RB}}\diag(\boldsymbol{\alpha}_{\mathrm{RB}})) (\bA_{\mathrm{R,UR}}^{\mathrm{T}} \diamond \bA_{\mathrm{R,RB}}^{\mathrm{H}}) \nonumber
		\\ &\mathop = \limits^{(b)} (\bI_K \otimes \bA_{\mathrm{B,RB}}) (\bLambda_{\mathrm{UR}}^{\mathrm{T}} \otimes \diag(\boldsymbol{\alpha}_{\mathrm{RB}})) (\bA_{\mathrm{R,UR}}^{\mathrm{T}} \diamond \bA_{\mathrm{R,RB}}^{\mathrm{H}}),
	\end{align}
	where $(a)$ and $(b)$ follow from the identities $\bM_1\bM_2 \diamond \bM_3\bM_4 = (\bM_1\otimes\bM_3)(\bM_2 \diamond \bM_4)$ and $\bM_1\bM_2 \otimes \bM_3\bM_4 = (\bM_1\otimes\bM_3)(\bM_2 \otimes \bM_4)$, respectively. Using the identity $\mathrm{vec}(\bM_1 \bM_2 \bM_3)=(\bM_3^{\mathrm{T}}\otimes \bM_1) \mathrm{vec}(\bM_2)$, $\bC$ in (\ref{cascade_reformulated}) can be vectorized~as
	\begin{align} \label{c_lemma_1_1}
		\bc &= ((\bA_{\mathrm{R,UR}}^{\mathrm{T}} \diamond \bA_{\mathrm{R,RB}}^{\mathrm{H}})^{\mathrm{T}} \otimes (\bI_K \otimes \bA_{\mathrm{B,RB}})) \nonumber \\& \quad \quad \cdot \mathrm{vec}(\bLambda_{\mathrm{UR}}^{\mathrm{T}} \otimes \diag(\boldsymbol{\alpha}_{\mathrm{RB}})) \nonumber
		\\ &= ((\bA_{\mathrm{R,UR}}^{\mathrm{T}} \diamond \bA_{\mathrm{R,RB}}^{\mathrm{H}})^{\mathrm{T}} \diamond \tilde{\bA}_{\mathrm{B,RB}}) (\boldsymbol{\alpha}_{\mathrm{UR}} \otimes \boldsymbol{\alpha}_{\mathrm{RB}}),
	\end{align}
	which completes the proof.

    \subsection{Proof of Lemma \ref{Lemma_digital}}
    In the considered system for hardware-limited task-based quantization, the scalar ADCs are modeled as non-subtractive uniform dithered quantizers, where the dithered signals are uniformly distributed over $(-\Delta/2, \Delta/2]$.
    Assuming that the inputs to the ADCs lie within their dynamic range with probability one, the outputs of the ADCs can be expressed as $\bB_{\bc} \by + \mathbf{e}$, where $\mathbf{e}$ denotes the quantization noise. Based on the results in \cite{Dithered_quantizer}, it is clear that $\mathbf{e}$ is uncorrelated with $\bB_{\bc} \by$, and the real and imaginary parts of each entry of  $\mathbf{e}$ are independent and have zero-mean and variance $\frac{\Delta^2}{6}$.
	Thus, given $\bB_{\bc}$, the optimal approach for estimating $\hat{\bc}$ is the linear MMSE estimator of $\tilde{\bc}= \bGamma_{\bc}\by$ from $\bB_{\bc}\by + \mathbf{e}$, given~by
    \begin{align} \label{Proof_Lemma2}
    	\bD^{\mathrm{o}}_{\bc}(\bB_{\bc}) &= \mathbb{E}[\tilde{\bc} (\bB_{\bc}\by + \mathbf{e})^{\mathrm{H}}] \cdot (\mathbb{E}[(\bB_{\bc}\by + \mathbf{e})(\bB_{\bc}\by + \mathbf{e})^{\mathrm{H}}])^{-1} \nonumber
    	\\ &= \mathbb{E}[\bGamma_{\bc} \by (\bB_{\bc} \by)^{\mathrm{H}}] \cdot (\mathbb{E}[(\bB_{\bc}\by)(\bB_{\bc}\by)^{\mathrm{H}}] + \mathbb{E}[\mathbf{e}\mathbf{e}^{\mathrm{H}}])^{-1} \nonumber
    	\\ &\mathop = \limits^{(a)} \bGamma_{\bc} \bSigma_{\by} \bB_{\bc}^{\mathrm{H}} \left(\bB_{\bc} \bSigma_{\by} \bB_{\bc}^{\mathrm{H}} + \frac{4\gamma_{\bc}^2}{3\tilde{\nu}_{\bc}^2}\bI_{G_{\bc}} \right)^{-1},
    \end{align}
    where $(a)$ follows from $\Delta = \frac{2 \gamma_{\bc}}{\tilde{\nu}_{\bc}}$. The resulting MSE in (\ref{Lemma_digital_MSE}) can be computed based on the result in (\ref{Proof_Lemma2}), which completes the proof.
	
    \subsection{Proof of Proposition \ref{Proposition_analog}}
    Based on the discussion in Section \ref{sec4_1}, we first define the ADC threshold $\gamma_{\bc}$, which is taken to be a multiple $\eta_{\bc}$ of the maximum standard deviation of its input, given by
    \begin{align} \label{Proof_Proposition1_gamma_c}
    	\gamma_{\bc}^2 &= \eta_{\bc}^2 \max_{g=1,\cdots,G_{\bc}} \mathbb{E} \left[ \vert [\bB_{\bc}\by]_g + \beta_g \vert^2 \right] \nonumber
    	\\ &=  \eta_{\bc}^2 \max_{g=1,\cdots,G_{\bc}} \mathbb{E} \left[ \vert [\bB_{\bc}\by]_g \vert^2 \right] + \eta_{\bc}^2\frac{2\gamma_{\bc}^2}{3 \tilde{\nu}_{\bc}} \nonumber
    	\\ &= \kappa_{\bc} \max_{g=1,\cdots,G_{\bc}} \mathbb{E} \left[ \vert [\bB_{\bc}\by]_g \vert^2 \right],
    \end{align}
	where $\beta_g$ is the dither signal, which is independent of $\by$ and has variance $\frac{\Delta^2}{6} = \frac{2 \gamma_{\bc}^2}{3 \tilde{\nu}_{\bc}}$ \cite{Dithered_quantizer}, and $\kappa_{\bc}=\eta_{\bc}^2 \left(1-\frac{2\eta_{\bc}^2}{3 \tilde{\nu}_{\bc}} \right)^{-1}.$
	Using (\ref{Proof_Proposition1_gamma_c}), given an analog combining matrix $\bB_{\bc}$, the achievable MSE defined in (\ref{Lemma_digital_MSE}) can be rewritten as
    \begin{align} \label{Proof_Proposition1_MSE}
    	&\mathrm{MSE}(\bB_{\bc}) = \trace \bigg(\bGamma_{\bc} \bSigma_{\by} \bGamma_{\bc}^{\mathrm{H}}- \bGamma_{\bc} \bSigma_{\by} \bB_{\bc}^{\mathrm{H}} \bigg(\bB_{\bc} \bSigma_{\by} \bB_{\bc}^{\mathrm{H}} \nonumber \\&+ \frac{4\kappa_{\bc}^2}{3\tilde{\nu}_{\bc}^2} \max_{g=1,\cdots,G_{\bc}} \mathbb{E}\left[ \vert [\bB_{\bc}\by]_{g} \vert^2 \right] \bI_{G_{\bc}} \bigg)^{-1} \bB_{\bc} \bSigma_{\by} \bGamma_{\bc}^{\mathrm{H}} \bigg).
    \end{align}

    To simplify (\ref{Proof_Proposition1_MSE}), we use the result in Lemma C.1 from  \cite{Shlezinger:2019}, which states that for any $\bB_{\bc}$, there exists a unitary matrix $\bU_{\bc} \in \mathbb{C}^{G_{\bc} \times G_{\bc}}$ such that $\mathrm{MSE}(\bB_{\bc}) \geq  \mathrm{MSE}(\bU_{\bc} \bB_{\bc})$. Furthermore, using Corollary 2.4 from \cite{Unitary_matrix}, we have
    \begin{align} \label{Proof_Proposition1_unitary}
    	\min_{\bU_{\bc}} \max_{g=1,\cdots,G_{\bc}} \left[\bU_{\bc} \bB_{\bc} \bSigma_{\by} \bB_{\bc}^{\mathrm{H}} \bU_{\bc}^{\mathrm{H}} \right]_{g,g} = \frac{1}{G_{\bc}} \trace \left( \bB_{\bc} \bSigma_{\by} \bB_{\bc}^{\mathrm{H}} \right),
    \end{align}
    where the $\bU_{\bc}$ that achieves this minimum value can be derived using Algorithm 2.2 in \cite{Unitary_matrix}.
    Combining the results from (\ref{Proof_Proposition1_MSE}) and (\ref{Proof_Proposition1_unitary}), and defining $\widetilde{\bB}_{\bc} = \bB_{\bc} \bSigma_{\by}^{\frac{1}{2}}$ and $\widetilde{\bGamma}_{\bc} = \bGamma_{\bc} \bSigma_{\by}^{\frac{1}{2}}$, the MSE minimization problem in (\ref{Proof_Proposition1_MSE}) simplifies to	
	\begin{align} \label{Proof_Proposition1_MSE_2}
		\max_{\widetilde{\bB}_{\bc}} \enspace \trace &\bigg( \widetilde{\bGamma}_{\bc} \widetilde{\bB}_{\bc}^{\mathrm{H}} \left(\widetilde{\bB}_{\bc} \widetilde{\bB}_{\bc}^{\mathrm{H}} +\frac{4\kappa_{\bc}}{3 \tilde{\nu}_{\bc}G_{\bc}}\trace(\widetilde{\bB}_{\bc}\widetilde{\bB}_{\bc}^{\mathrm{H}}) \bI_{G_{\bc}} \right)^{-1} \nonumber
		\\ & \quad \times \widetilde{\bB}_{\bc} \widetilde{\bGamma}_{\bc}^{\mathrm{H}} \bigg).
	\end{align}
	
	To proceed further, we use the fact that the MSE in (\ref{Proof_Proposition1_MSE_2}) remains unchanged when $\widetilde{\bB}_{\bc}$ is replaced with $\alpha\bU \widetilde{\bB}_{\bc} $, where $\bU$ is a unitary matrix and $\alpha > 0$, allowing us to set $\trace(\widetilde{\bB}_{\bc} \widetilde{\bB}_{\bc}^{\mathrm{H}})=1$.
	Letting $\widetilde{\bB}_{\bc} = \bLambda_{\bc} \bV_{\bc}^{\mathrm{H}}$ with diagonal matrix $\bLambda_{\bc} \in \mathbb{C}^{G_{\bc} \times NT\tau}$ and unitary matrix $\bV_{\bc} \in \mathbb{C}^{NT\tau \times NT\tau}$, the problem in (\ref{Proof_Proposition1_MSE_2}) can be reformulated as
    \begin{align} \label{Proof_Proposition1_MSE_3}
    	\max_{\bLambda_{\bc}, \bV_{\bc}} \enspace & \trace  \left( \widetilde{\bGamma}_{\bc}^{\mathrm{H}} \widetilde{\bGamma}_{\bc} \bV_{\bc} \bLambda_{\bc}^{\mathrm{H}} \left( \bLambda_{\bc}\bLambda_{\bc}^{\mathrm{H}} + \frac{4\kappa_{\bc}}{3 \tilde{\nu}_{\bc}G_{\bc}}\bI_{G_{\bc}} \right)^{-1} \bLambda_{\bc}\bV_{\bc}^{\mathrm{H}}   \right) \nonumber
    	\\ \mbox{s.t. } \enspace & \trace\left( \bLambda_{\bc} \bLambda_{\bc}^{\mathrm{H}}\right)=1.
    \end{align}
	
    Finally, let $\widetilde{\bLambda}_{\bc}=\bLambda_{\bc}^{\mathrm{H}} \left( \bLambda_{\bc}\bLambda_{\bc}^{\mathrm{H}} + \frac{4\kappa_{\bc}}{3 \tilde{\nu}_{\bc}G_{\bc}}\bI_{G_{\bc}} \right)^{-1} \bLambda_{\bc}$, which is a diagonal matrix. Based on Theorem II.1 in \cite{Lasserre:1995}, it can be shown that the optimal $\bV_{\bc}$ for the problem in (\ref{Proof_Proposition1_MSE_3}) is the matrix of right singular vectors corresponding to $\widetilde{\bLambda}_{\bc}$ when its entries are arranged in descending order.
    Based on this result, it is clear that the objective function in (\ref{Proof_Proposition1_MSE_3}) is concave with respect to $\{([\bLambda_{\bc}]_{g,g})^2\}_{g=1}^{G_{\bc}}$, and the resulting solution satisfying the Karush–Kuhn–Tucker (KKT) conditions leads to the expression given in (\ref{Lambda_c}), where we implicitly assume $NT\tau > G_{\bc}$. Additional details can be found in \cite{Shlezinger:2019}.
    Thus, combining the results discussed so far, the optimal analog combining matrix minimizing the MSE is given by $\bB^{\mathrm{o}}_{\bc} = \bU_{\bc} \bLambda_{\bc} \bV_{\bc}^{\mathrm{H}} \bSigma_{\by}^{-\frac{1}{2}}$, and the corresponding ADC threshold is $\gamma_{\bc}^2 = \frac{\kappa_{\bc}}{G_{\bc}} \trace \left( \bLambda_{\bc} \bLambda_{\bc}^{\mathrm{H}} \right) = \frac{\kappa_{\bc}}{G_{\bc}}$, which completes the~proof.
    
	\subsection{Proof of Lemma \ref{Lemma_W_c}}
	To analyze the rank of $\bW_{\bc}=(\bA_{\mathrm{R,UR}}^{\mathrm{T}} \diamond \bA_{\mathrm{R,RB}}^{\mathrm{H}})^{\mathrm{T}} \diamond \tilde{\bA}_{\mathrm{B,RB}}$, we first investigate the rank of $(\bA_{\mathrm{R,UR}}^{\mathrm{T}} \diamond \bA_{\mathrm{R,RB}}^{\mathrm{H}})^{\mathrm{T}}$, which is the cascaded array response matrix at the RIS \cite{Pan:2022, Chen:2023}, whose $s$-th column is given by
	\begin{align} \label{Cascaded_array}
		&\left[(\bA_{\mathrm{R,UR}}^{\mathrm{T}} \diamond \bA_{\mathrm{R,RB}}^{\mathrm{H}})_{s,:} \right]^{\mathrm{T}} \nonumber \\&= [\ba_{\mathrm{R}}^{\mathrm{T}}(\theta_{\mathrm{UR},k,i_k}^{\mathrm{Azi}}, \theta_{\mathrm{UR},k,i_k}^{\mathrm{Ele}}) \diamond \ba_{\mathrm{R}}^{\mathrm{H}}(\theta_{\mathrm{RB},j}^{\mathrm{Azi}}, \theta_{\mathrm{RB},j}^{\mathrm{Ele}})]^{\mathrm{T}} \nonumber \\ &= \ba_{\mathrm{R}}(\theta_{\mathrm{UR},k,i_k}^{\mathrm{Azi}} - \theta_{\mathrm{RB},j}^{\mathrm{Azi}}, \theta_{\mathrm{UR},k,i_k}^{\mathrm{Ele}} - \theta_{\mathrm{RB},j}^{\mathrm{Ele}}),
	\end{align}
	where $i_k=i-\sum_{r=1}^{k-1} M_{\mathrm{UR},r}$ with $i = \lceil \frac{s}{M_{\mathrm{RB}}} \rceil$ when $i$ is within the range $\sum_{r=1}^{k-1} M_{\mathrm{UR},r} < i \leq\sum_{r=1}^{k} M_{\mathrm{UR},r}$, and $j=((s-1) \mbox{ mod } M_{\mathrm{RB}})+1$.
	From (\ref{Cascaded_array}), when the second condition in Lemma \ref{Lemma_W_c} is satisfied, all columns in $(\bA_{\mathrm{R,UR}}^{\mathrm{T}} \diamond \bA_{\mathrm{R,RB}}^{\mathrm{H}})^{\mathrm{T}}$ are linearly independent, implying that $(\bA_{\mathrm{R,UR}}^{\mathrm{T}} \diamond \bA_{\mathrm{R,RB}}^{\mathrm{H}})^{\mathrm{T}}$ has full column rank $M_{\mathrm{RB}}M_{\mathrm{UR}}$ when $L \geq M_{\mathrm{RB}}M_{\mathrm{UR}}$. To proceed further, we introduce the following lemma regarding the rank of a Khatri-Rao~product.
	\begin{lemma} \label{Lemma_Khatri}
		If $\mathrm{rank}((\bA_{\mathrm{R,UR}}^{\mathrm{T}} \diamond \bA_{\mathrm{R,RB}}^{\mathrm{H}})^{\mathrm{T}}) + \mathrm{rank}(\tilde{\bA}_{\mathrm{B,RB}} ) \geq M_{\mathrm{RB}}M_{\mathrm{UR}} + 1$, $\bW_{\bc}$ has full column rank $M_{\mathrm{RB}}M_{\mathrm{UR}}$.
		\begin{proof}
			The proof is based on Lemma 1 in \cite{Sidiropoulos:2000}.
		\end{proof}
	\end{lemma}
	From Lemma \ref{Lemma_Khatri}, since $\mathrm{rank}(\tilde{\bA}_{\mathrm{B,RB}}) \geq 1$ always holds as long as it is not an all-zero matrix, $\bW_{\bc}$ has full column rank $M_{\mathrm{RB}} M_{\mathrm{UR}}$ when $(\bA_{\mathrm{R,UR}}^{\mathrm{T}} \diamond \bA_{\mathrm{R,RB}}^{\mathrm{H}})^{\mathrm{T}}$ is a full column rank matrix, which completes the proof.
    
	\subsection{Proof of Lemma \ref{Lemma_training_overhead}}
	Under the considered conditions, the rank of $\bGamma_{\bc}$ in (\ref{Gamma_c}) depends on the rank of $\mathbb{E}[\bc \by^{\mathrm{H}}]$, which is upper bounded by $\min(M_{\mathrm{RB}}M_{\mathrm{UR}}, \mathrm{rank}(\bar{\bS}\bW_{\bc}) )$. Defining $\bA_{\bc} = (\bA_{\mathrm{R,UR}}^{\mathrm{T}} \diamond \bA_{\mathrm{R,RB}}^{\mathrm{H}})^{\mathrm{T}}$, $\bar{\bS}\bW_{\bc}$ can be reformulated as
	\begin{align} \label{Lemma_SWc}
		\bar{\bS}\bW_{\bc} &= (\bS^{\mathrm{T}} \otimes \bX_{\mathrm{C}}^{\mathrm{T}} \otimes \bI_N) (\bA_{\bc} \diamond \tilde{\bA}_{\mathrm{B,RB}}) \nonumber
		\\ &= (\bS^{\mathrm{T}} \bA_{\bc}) \diamond ( (\bX_{\mathrm{C}}^{\mathrm{T}} \otimes \bI_N) \tilde{\bA}_{\mathrm{B,RB}}).
	\end{align}
	Let $\bR = (\bX_{\mathrm{C}}^{\mathrm{T}} \otimes \bI_N) \tilde{\bA}_{\mathrm{B,RB}}$. Now, to derive the tightest conditions for $T$, we consider the case where $T\leq M_{\mathrm{RB}}M_{\mathrm{UR}}$ and $KM_{\mathrm{RB}} \leq N\tau$, leading to $\mathrm{rank}(\bS \bA_{\bc})=T$ and $\mathrm{rank}(\bR)=KM_{\mathrm{RB}}$.
	The rank of $\bar{\bS}\bW_{\bc}$ is then upper bounded~by
	\begin{align} \label{Lemma_rank_SWc}
		&\mathrm{rank}(\bar{\bS}\bW_{\bc}) \nonumber
		\\&= \mathrm{rank}( ((\bS^{\mathrm{T}} \bA_{\bc}) \diamond \bR)^{\mathrm{H}} ((\bS^{\mathrm{T}} \bA_{\bc}) \diamond \bR)) \nonumber
		\\ & \mathop = \limits^{(a)} \mathrm{rank}((\bS^{\mathrm{T}} \bA_{\bc})^{\mathrm{H}}(\bS^{\mathrm{T}} \bA_{\bc})) \odot (\bR^{\mathrm{H}} \bR)) \nonumber
		\\ & \mathop \leq \limits^{(b)} KTM_{\mathrm{RB}},
	\end{align}
	where $(a)$ holds due to $(\bM_1 \diamond \bM_2)^{\mathrm{H}}(\bM_1 \diamond \bM_2) = (\bM_1^{\mathrm{H}} \bM_1)\odot (\bM_2^{\mathrm{H}}\bM_2)$, and $(b)$ follows from $\mathrm{rank}(\bM_1 \odot \bM_2) \leq \mathrm{rank}(\bM_1) \cdot \mathrm{rank}(\bM_2)$.
	From (\ref{Lemma_rank_SWc}), it is clear that $\mathrm{rank}(\bar{\bS}\bW_{\bc}) \leq \min(KTM_{\mathrm{RB}},$ $NT\tau, M_{\mathrm{RB}}M_{\mathrm{UR}})$. Thus, the necessary conditions for $\bGamma_{\bc}$ to have its maximum rank are: 1) $NT\tau \geq M_{\mathrm{RB}}M_{\mathrm{UR}}$, and 2) $KT\geq M_{\mathrm{UR}}$, which completes the~proof.
}

\bibliographystyle{IEEEtran}
\bibliography{refs_all}
\end{document}